\documentclass[%
 aip,
sor,
 amsmath,amssymb,
 reprint,%
]{revtex4-1}

\usepackage{xcolor}
\usepackage{bm}
\usepackage{hyperref}
\usepackage{graphicx}
\usepackage{subfigure}
\usepackage{siunitx}
\usepackage{natbib}

\definecolor{Red}{rgb}{1,0.0,0.0}
\renewcommand{\vec}[1]{\bm{#1}}
\newcommand{\tsor}[1]{\bm{#1}}

\usepackage{amsmath}

\begin{document}
%
\title{A constitutive model for simple shear of dense frictional suspensions}
\author{Abhinendra Singh}
\email[]{asingh@ccny.cuny.edu}
\affiliation{Benjamin Levich Institute, CUNY City College of New York, New York, NY 10031, USA.}
\author{Romain Mari}
\affiliation{Department of Applied Mathematics and Theoretical Physics, Centre for Mathematical Sciences, University of Cambridge, Cambridge CB3 0WA, United Kingdom.}
\author{Morton M. Denn}
\affiliation{Benjamin Levich Institute, CUNY City College of New York, New York, NY 10031, USA.}
\affiliation{Department of Chemical Engineering, CUNY City College of New York, New York, NY 10031.}
\author{Jeffrey F. Morris}
\affiliation{Benjamin Levich Institute, CUNY City College of New York, New York, NY 10031, USA.}
\affiliation{Department of Chemical Engineering, CUNY City College of New York, New York, NY 10031.}

\date{\today}
\begin{abstract}
 Discrete particle simulations are used to study the shear rheology of dense, stabilized, frictional particulate suspensions in a viscous liquid, 
 toward development of a constitutive model for steady shear flows at arbitrary stress.  
 These suspensions undergo increasingly strong continuous shear thickening (CST) as solid volume fraction $\phi$ increases
 above a critical volume fraction, and discontinuous shear thickening (DST) is observed for a range of $\phi$.
  When studied at controlled stress, the DST behavior is associated with non-monotonic flow curves of the steady-state stress as a function of shear rate.
  Recent studies have related shear thickening to a transition between mostly lubricated to predominantly frictional 
  contacts with the increase in stress.
   In this study, the behavior is simulated over a wide range of the dimensionless parameters $(\phi,\tilde{\sigma}$, and $\mu)$,
    with $\tilde{\sigma} = \sigma/\sigma_0$ the dimensionless shear stress and $\mu$ the coefficient of interparticle friction:
    the dimensional stress is $\sigma$, and $\sigma_0 \propto F_0/ a^2$, where 
    $F_0$ is the magnitude of repulsive force at contact and $a$ is the particle radius.
   The data have been used to populate the model of the lubricated-to-frictional rheology of Wyart and Cates [Phys. Rev. Lett.{\bf 112}, 098302 (2014)], which is based on the concept of two viscosity divergences or \textquotedblleft jamming\textquotedblright\ points at volume
   fraction $\phi_{\rm J}^0 = \phi_{\rm rcp}$ (random close packing) for the low-stress lubricated state, and at 
 $\phi_{\rm J} (\mu) < \phi_{\rm J}^0$ for any nonzero $\mu$ in the frictional state; a generalization provides the normal stress response as well as the shear stress.
A flow state map of this material is developed based on the simulation results.  
At low stress and/or intermediate $\phi$, the system exhibits CST, and DST appears
at volume fractions below but approaching the frictional jamming point. 
 For $\phi< \phi_{\rm J}^\mu$, DST is associated with a material transition from one stress-independent rheology to another,
 while for $\phi> \phi_{\rm J}^\mu$, the system exhibits DST to shear jamming as the stress increases.
\end{abstract}
\pacs{}
\maketitle
\section{Introduction}
Dense non-Brownian suspensions of rigid particles exhibit diverse rheological behavior,
including yielding, shear thinning or shear thickening, normal stress differences, particle migration and
shear jamming~\citep{mewis_colloidal_2011, guazzelli_physical_2011, Denn_2014, Peters_2016}.
Shear thickening of dense suspensions is a phenomenon in which, for a range of applied shear stress,
the apparent viscosity increases, sometimes by an order of magnitude or more \cite{Metzner_1958,Barnes_1989, Brown_2014}.
Strong continuous shear thickening (CST) is observed in concentrated suspensions, where 
the viscosity increases continuously with increase in shear rate. 
At volume fractions above a critical value an abrupt increase in viscosity
may be observed and is termed discontinuous shear thickening (DST).

Recent experimental~\cite{Fernandez_2013,Pan2015s,Guy_2015,Clavaud_2017,Comtet_2017}
and computational~\cite{Seto_2013a,Heussinger_2013,Fernandez_2013,Mari_2014,Mari_2015,Ness2016} work has demonstrated
that shear thickening (both CST and DST) can arise due to frictional particle-particle contacts.
In a suspension, a repulsive force is often present due to steric (e.g. due to adsorbed polymer) or electrostatic stabilization.
When the shear forces acting to bring a pair of particles into contact exceed the repulsive force threshold,
only fluid mechanical forces are available to keep the particle surfaces apart, and Melrose and Ball~\cite{Melrose_1995}
have shown that the lubrication film can go to arbitrarily small values. 
We therefore assume the lubrication film can break, allowing for contact interactions of both 
normal and tangential (frictional) form, which most simply can be seen as representative of direct surface 
contacts due to roughness, but could also be a result of other phenomena, e.g. polymer brush interactions 
\cite{Fernandez_2013}.
Thus the repulsive force $F_0$ gives rise to a stress scale $\sigma_0 = F_0/6\pi a^2$ for particles of radius $a$,
which marks a crossover from lubricated (frictionless) contacts between particles to direct, frictional contacts.
In an idealized model where stabilization forces themselves do not 
contribute to the stress and only act as a switch for friction (a 
realization of which is possible in simulations, see the discussion of the ``critical load model'' 
\cite{Mari_2014}), at low stress ($\sigma \ll \sigma_0$) the particle interactions are lubricated (frictionless)
so that the system is rate-independent with the viscosity diverging at the frictionless jamming point $\phi_{\rm J}^0$, 
corresponding to random close packing of $\phi_{\rm rcp} \approx 0.64$ for monodisperse spheres.
On the other hand, in the shear-thickened state (for $\sigma \gg \sigma_0$) almost all contacts are frictional,
the viscosity is again rate-independent but diverges at a volume fraction $\phi_{\rm J}^\mu < \phi_{\rm J}^0$,
where $\phi_{\rm J}^\mu = \phi_{\rm J}(\mu)$ is used as shorthand to denote the dependence of this jamming fraction on the interparticle friction coefficient 
$\mu$~\cite{Silbert_2010,Otsuki_2011,Ciamarra_2011,Chialvo2012bridging,Mari_2014}.
With increase in $\sigma$, the crossover between these two states results in the shear thickening behavior.
At large $\phi$, a finite-range stabilizing force becomes one of 
the major sources of stress at low stresses, which leads to a 
shear-thinning rheology. In this work, we focus on developing a 
constitutive model for the shear-thickening part of the flow curves. Our 
proposed model does not include the physics behind the shear-thinning at 
low stresses, but as this is due to an unrelated microscopic mechanism with 
its own set of parameters, our idealized model could in principle be 
augmented to include the relevant low stress physics and display both 
shear thinning and shear thickening.

\begin{figure}
\centering
\includegraphics[width=.48\textwidth]{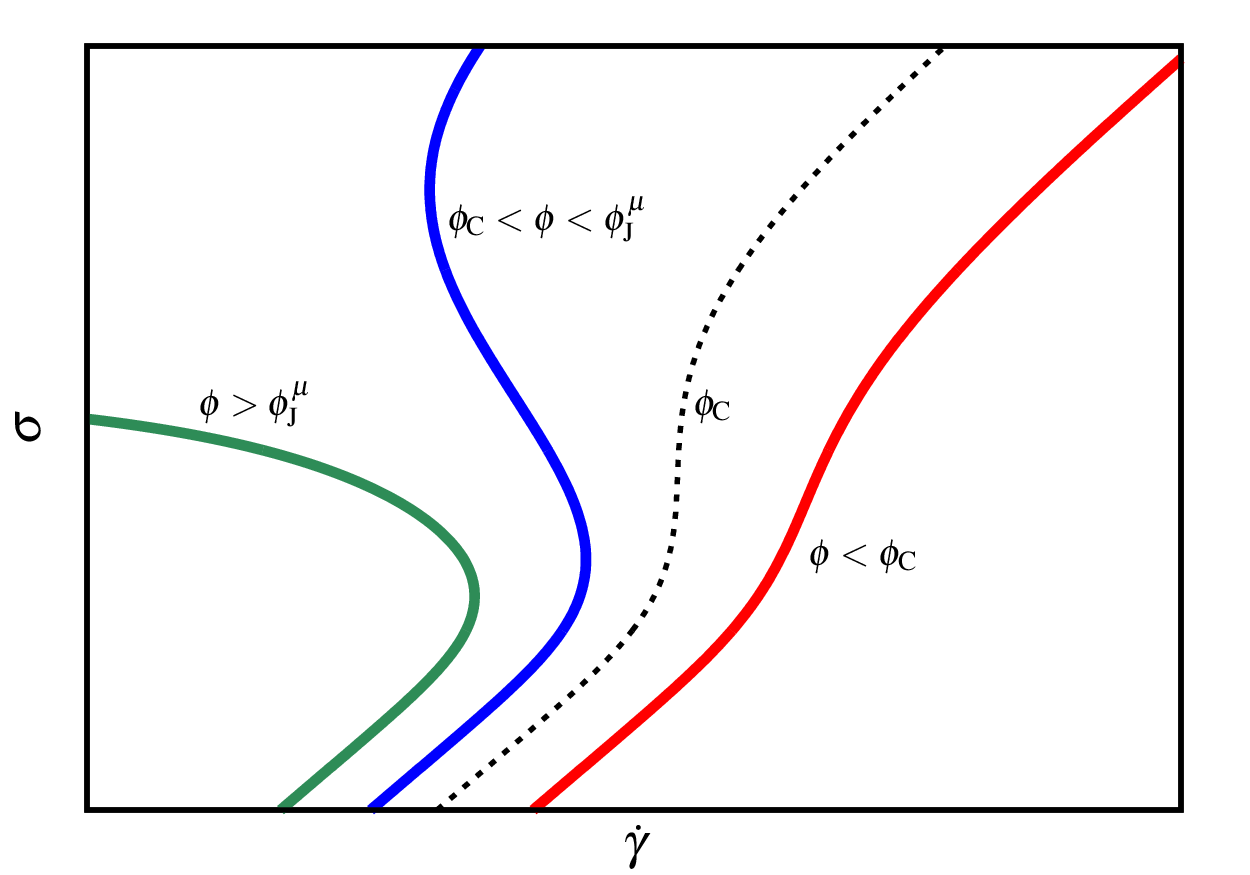}
\caption{ 
Sketch of the relation between the shear stress $\sigma$
and the shear rate $\dot{\gamma}$  in a shear thickening
suspension, with increasing volume fraction $\phi$ from left to right.
For $\phi<\phi_{\rm C}$ (dotted line), $\sigma(\dot{\gamma})$ is monotonic and the
shear thickening is continuous. $\phi = \phi_{\rm C}$ corresponds to the ``critical volume fraction"
at which ${\sigma}(\dot{\gamma})$ has a point where it shows infinite slope.
In the range $\phi_{\rm C}<\phi<\phi_{\rm J}^\mu$, the flow curve becomes non-monotonic
and suspension undergoes discontinuous shear thickening.
For $\phi > \phi_{\rm J}^\mu$ , the backward bending branch hits the vertical axis.
This means that the suspension can flow only at small shear stress.
}
\label{fig:sketch}
\end{figure}

Along this line of thought, aspects of which had been developed in Bashkirtseva {\it et al.}~\cite{Bashkirtseva_2009}, 
\citet{Wyart_2014} (WC) have shown that under rather broad conditions
a viscosity increasing with shear stress by interpolating between two rate-independent asymptotic rheologies
can lead to three forms of shear stress curves as a function of shear rate (cf. Fig.~\ref{fig:sketch}),
depending on the viscosity difference between the two states representative of unthickened and thickened suspensions.
When this difference is small, i.e. for $\phi \ll \phi_{\rm J}^\mu$, the shear thickening is continuous, with a
monotonic relation between shear stress and rate.
For a large enough but finite viscosity contrast,  i.e. when $\phi_{\mathrm{C}}<\phi<\phi_{\rm J}^\mu$,
the flow curve $\sigma(\dot{\gamma})$ becomes non-monotonic, S-shaped, and the thickening becomes discontinuous.
Finally, when the thickened branch is actually jammed, i.e. for $\phi \geq \phi_{\rm J}^\mu$,
the system can flow only for low stresses, while at high stresses frictional contacts cause the system to shear jam,
and in this case the flow curve at large stress tends toward a zero shear rate state.
Non-monotonic  flow curves have been reported under controlled stress conditions in several subsequent experimental
and simulation studies~\cite{Neuville_2012,Mari_2015,Pan2015s,Hermes_2016}.

\citet{Wyart_2014} have also proposed a rheological model for shear thickening exhibiting the features noted above.
This model is based on an interpolation between two diverging stress-independent rheologies,
where the interpolation depends on a unique microscopic state parameter identified as the ``fraction of frictional contacts'', $f$.
Wyart and Cates described $f$ as a function of $\Pi$, the particle pressure, but in standard rheometric experiments, where $\phi$ is fixed, 
$\Pi$ and $\sigma$ are directly related~\citep{Morris_1999}.
 Since $\sigma$ is more readily controlled, we find it more convenient to consider $f(\tilde{\sigma})$,
 where $\tilde{\sigma} = \sigma/\sigma_0$.  
 In rate-controlled simulations, $f$ has been found to be a 
function of $\tilde{\sigma}$ \cite{Mari_2014}, while in 
pressure-controlled simulations, $f$ is not found to be a unique function of $\Pi$ \cite{Dong_2017}.
The quantity $f$ serves, in essence, as an order parameter for the shear thickening transition,
assuming low values in the low viscosity state under small stress,
 and asymptoting to $f \approx 1$ in the large viscosity state at large stress.

In this article, we explore the WC model by comparison to extensive numerical simulations we have performed by 
varying systematically the volume fraction and friction coefficient of the microscopic model, and exploring the resulting stress (or rate) dependence.
Informed by the numerical results, we then introduce empirical expressions
for the relations between the parameters $\mu$, $\phi$, the order parameter $f$, and the steady-state stress.
We find that upon increasing the friction coefficient, the frictional jamming point $\phi_{\rm J}^\mu$ decreases,
but the power-law exponent of divergence for both shear and normal stresses, $\sim (\phi_{\mathrm{J}}-\phi)^{-\beta}$ remains well-described by $\beta = 2$.
We show that the WC model is highly effective at predicting
the shear stress for a wide range of $\mu$ and $\phi$.
An extension following the same model structure provides predictions for the normal stress response to allow for a complete description of 
the viscometric functions in steady simple-shear flow of shear thickening suspensions.
%


\section{Simulation methods}

We simulate an assembly of inertialess frictional spheres
immersed in a Newtonian fluid under an imposed shear stress $\sigma$,
giving rise to an imposed velocity field $\vec{v} = \dot\gamma(t)\hat{\vec{v}}(\bm{x}) = \dot\gamma(t)(x_2, 0, 0)$.
We use Lees-Edwards periodic boundary conditions with $N=500$ particles in a unit cell.
To avoid ordering, we use bidisperse particles, with radii $a$ and $1.4a$ mixed at equal volume fractions.
The particles interact through short-range hydrodynamic forces (lubrication), a short-ranged repulsive force
and frictional contacts; this simulation model that has been shown to reproduce accurately
many features of 
the experimentally measured rheology for dense shear-thickening suspensions~\cite{Seto_2013a,Mari_2014}, 
although discrepancies have been noted~\cite{mari_discontinuous_2015} in the small first normal stress difference
relative to some experimental observations~\cite{Cwalina_2014}.

The equation of motion for $N$ spheres is
the $6N$-dimensional force/torque balance between hydrodynamic ($\vec{F}_{\mathrm{H}}$),
repulsive ($\vec{F}_{\mathrm{R}}$), and contact ($\vec{F}_{\mathrm{C}}$) interactions,
\begin{equation}
  \vec{0} = \vec{F}_{\mathrm{H}}(\vec{X},\vec{U}) + \vec{F}_{\mathrm{C}}(\vec{X}) + \vec{F}_{\mathrm{R}}(\vec{X}), 
  \label{eq:force_balance}
\end{equation}
where the particle positions are denoted by $\vec{X}$ and their velocities/angular velocities by $\vec{U}$.
$\vec{F}_{\mathrm{R}}$ is a conservative force and can be determined based on the positions $\vec{X}$ of the particles, while
the calculation of the tangential component of the contact force $\vec{F}_{\mathrm{C}}$ is more involved as it depends
on the deformation history of the contact.

We make the translational velocities dimensionless with $\dot{\gamma}a$ and the shear rate and rotation rates by $\dot{\gamma}$.  
Decomposing the dimensionless velocity as $\hat{\vec{v}}(\vec{r})=\hat{\vec{\omega}}\times \vec{r} + \hat{\tsor{e}}\cdot\vec{r}$ in rotational
$\hat{\vec{\omega}} = (0,0,-1/2)$ and extensional $\hat{\tsor{e}}_{12}=\hat{\tsor{e}}_{21}=1/2$ parts,
the hydrodynamic force and torque vector takes the form
\begin{equation}
  \vec{F}_{\mathrm{H}}(\vec{X},\vec{U}) =
  -\tsor{R}_{\mathrm{FU}}(\vec{X}) \cdot \bigl(\vec{U}-\dot\gamma\hat{\vec{U}}^{\infty} \bigr)
  + \dot\gamma\tsor{R}_{\mathrm{FE}}(\vec{X}):\hat{\tsor{E}}, \label{eq:hydro_force}
\end{equation}
with $\hat{\vec{U}}^{\infty} = (\hat{\vec{v}}(y_1), \dots, \hat{\vec{v}}(y_N), \hat{\vec{\omega}}(y_1), \dots, \hat{\vec{\omega}}(y_N))$
and $\hat{\tsor{E}} = (\hat{\tsor{e}}(y_1), \dots, \hat{\tsor{e}}(y_N))$.
The position dependent resistance tensors
$\tsor{R}_{\mathrm{FU}}$ and $\tsor{R}_{\mathrm{FE}}$ include the
 ``squeeze'', ``shear'' and ``pump'' modes of pairwise lubrication~\citep{Ball_1997}, as well as one-body Stokes drag.
The occurrence of contacts between particles due, for example, to surface roughness is mimicked
by a regularization of the resistance divergence at vanishing interparticle gap
$h_{ij} = 2(r_{ij}-a_i-a_j)/(a_i+a_j)$: the ``squeeze'' mode resistance is proportional to $ 1/(h+\delta)$, while
the ``shear'' and ``pump'' mode resistances are proportional to $ \log(h+\delta)$~\citep{Mari_2014}.
Here we take $\delta=10^{-3}$.

The electrostatic repulsion force
is taken to represent a simple electrostatic double layer interaction between particles, with the force decaying exponentially with the interparticle surface separation $h$ as 
$|F_R|  = F_0 \exp(-h/\lambda)$, with a  Debye length $\lambda$.

Contacts are modeled by linear springs and dashpots. Tangential and normal components of the contact force
$F_C^{(ij)}$ between two particles satisfy Coulomb's friction
law  $|F_{C,t}^{(ij)}| \le \mu|F_{C,n}^{(ij)}|$, where $\mu$ is the interparticle friction coefficient.
Some softness is allowed in the contact, but the spring stiffnesses are taken such that the largest particle overlaps do not exceed \SI{3}{\percent}
of the particle radius during the simulation.

The equation of motion~\eqref{eq:force_balance} is solved under the
constraint of flow at constant shear stress $\sigma$.
At any time, the shear stress in the suspension is given by 
\begin{equation}
  \label{eq:stress}
  \sigma = \Sigma_{12} = \dot\gamma \eta_0\biggl(1+\frac{5}{2}\phi\biggr)
  + \dot\gamma \eta_{\mathrm{H}}+\sigma_{\mathrm{R}}+ \sigma_{\mathrm{C}}
\end{equation}
where $\eta_0$ is the suspending fluid viscosity,
$\eta_{\mathrm{H}} \dot{\gamma} = \dot{\gamma} V^{-1} \bigl\{ (\bm{R}_{\mathrm{SE}}
-\bm{R}_{\mathrm{SU}}\cdot\bm{R}_{\mathrm{FU}}^{-1}
\cdot\bm{R}_{\mathrm{FE}} ) :\hat{\bm{E}}^{\infty}\bigr\}_{12} $ is the contribution of 
hydrodynamic interactions to the stress, 
and
$\sigma_{\mathrm{R,C}} = V^{-1}\bigl\{ \bm{X}\bm{F}_{\mathrm{R,C}}
-\bm{R}_{\mathrm{SU}}\cdot\bm{R}_{\mathrm{FU}}^{-1}\cdot\bm{F}_{\mathrm{R,C}}
\bigr\}_{12}$, where $\bm{R}_{\mathrm{SU}}$
and $\bm{R}_{\mathrm{SE}}$ are resistance matrices
giving the lubrication stresses from the particles velocities
and resistance to deformation, respectively~\citep{Jeffrey_1992,Mari_2014},
and $V$ is the volume of the simulation box. Note that the resistance tensors are proportional to the 
suspending fluid viscosity. 
At fixed shear stress $\sigma$ the shear rate $\dot\gamma$ is the dependent variable
that is to be determined at each time step by~\citep{Mari_2015}
\begin{equation}
  \dot{\gamma} = \frac{\sigma - \sigma_{\mathrm{R}}-
    \sigma_{\mathrm{C}}}{\eta_0\Bigl(1+2.5\phi \Bigr) + \eta_{\mathrm{H}}}.
\end{equation}
The full solution of the equation of motion \eqref{eq:force_balance}
under the constraint of fixed stress \eqref{eq:stress}
is thus the velocity~\citep{Mari_2015}
\begin{equation}
 \bm{U}  = \dot{\gamma}\hat{\bm{U}}^{\infty}
  +
  \bm{R}_{\mathrm{FU}}^{-1}\cdot
  \bigl(
  \dot\gamma
  \bm{R}_{\mathrm{FE}}:\hat{\bm{E}}^{\infty}
  + \bm{F}_{\mathrm{R}}
  + \bm{F}_{\mathrm{C}}
\bigr).
\end{equation}

From these velocities, the positions are updated at each time step.
Lastly, the unit scales
are $\dot{\gamma}_0 \equiv F_{\rm 0}/{6\pi \eta_0 a^2}$ 
for the strain rate and $\sigma_0 \equiv \eta_0 \dot{\gamma}_0 = \frac{F_{\rm 0}}{{6\pi a^2}}$ for the stress.
In the rest of the paper, we use scaled stress defined as $\tilde{\sigma} = \sigma/\sigma_0$.

\section{Model}
In this section we present simulation results for values of friction coefficient
 $\mu =0.1, 0.2, 0.5, 1, 5$, and $10$.  As the friction coefficient determines $\phi_{\rm J}^\mu$, the results allow an exploration of 
 the effect of the separation between $\phi_{\rm J}^\mu$ and $\phi_{\rm J}^0$ on the rheology.
 As previously shown~\cite{Silbert_2010,Otsuki_2011,Chialvo2012bridging}, $0.1 \le \mu \le 1.0$ corresponds to the 
 moderate friction limit, where the friction coefficient affects the jamming point $\phi_{\rm J}^\mu$; $\mu >1$ corresponds
 to large friction, and we find that $\phi_{\rm J}^\mu$ saturates rapidly with $\mu>1$.

 In the following, the shear stress $\sigma$, particle pressure $\Pi$, and normal stress differences $N_1$ and $N_2$ are defined as $\sigma \equiv \Sigma_{12}$,
 $\Pi \equiv -(\Sigma_{11}+\Sigma_{22}+\Sigma_{33})/3$, $N_1 \equiv (\Sigma_{11}-\Sigma_{22})$, and $N_2 \equiv (\Sigma_{22}-\Sigma_{33})$, respectively. 
 All the stress components are non-dimensionalized by $\eta_0\dot{\gamma}_0$. 
 The dimensionless particle pressure $\Pi/\eta_0\dot{\gamma} \equiv \eta_{\rm n}$ represents
  the \textquotedblleft normal stress viscosity\textquotedblright\ \citep{Morris_1999}.
%
 
 The basic assumption of the model \cite{Wyart_2014} is that 
 $\eta_{\rm r}$, $\Pi/\eta_0\dot{\gamma}_0$ and $N_2/\eta_0\dot{\gamma}_0$ are in distinct stress-independent states both at low $(\tilde{\sigma} \ll 1)$ and high $(\tilde{\sigma} \gg 1)$ stress. 
 Here we model the $\phi$ dependence of viscosity and normal stresses as $ (\phi_{\rm J} - \phi)^{-2}$ and $\phi^2(\phi_{\rm J}-\phi)^{-2}$, respectively.
 %
 The forms are consistent with the proposed correlations for constant volume \citep{Morris_1999, Mills_2009} as well as constant pressure conditions \citep{Boyer_2011}.  

  The viscosity, particle pressure, and second normal stress difference as functions of $\phi$ in low and high stress states can be expressed as
   \begin{subequations}\label{eq:eta_P_N2_phi}
    \begin{equation}\label{eq:etaL_phi}
    \eta_{\rm r}^{\rm L} (\phi) = \alpha^0 (\phi_{\rm J}^0-\phi)^{-2}
   \end{equation}
   \begin{equation}
   \label{eq:etaH_phi}
    \eta_{\rm r}^{\rm H}(\phi,\mu)= \alpha^\mu (\phi_{\rm J}^\mu-\phi)^{-2}~,
   \end{equation}
   
    \begin{equation}\label{eq:P_phi_mu}
 \frac{\Pi^{\rm L}}{\eta_0\dot{\gamma}} (\phi)= \beta^0 {\phi}^2 (\phi_{\rm J}^0-\phi)^{-2}~,
 \end{equation}
 \begin{equation}
 \frac{\Pi^{\rm H}}{\eta_0\dot{\gamma}}(\phi,\mu)  = \beta^{\mu} {\phi}^2 (\phi_{\rm J}^{\mu}-\phi)^{-2}~,
 \end{equation}
 
  \begin{equation}\label{eq:N2_phi_mu}
  \frac{N_2^{\rm L}}{\eta_0\dot{\gamma}} (\phi) = K_2^0 {\phi}^2 (\phi_{\rm J}^0-\phi)^{-2}~,
 \end{equation}
 \begin{equation}
 N_2^{\rm H}/{\eta_0\dot{\gamma}}(\phi,\mu)= K_2^{\mu} {\phi}^2 (\phi_{\rm J}^{\mu}-\phi)^{-2}~,
 \end{equation}
   \end{subequations}
   where $\alpha^{0,\mu}$, $\beta^{0,\mu}$, $K_2^{0,\mu}$ are constant coefficients.
Recall that  $\phi_{\rm J}^0$ and $\phi_{\rm J}^\mu$ denote the jamming volume fraction for $\mu = 0$ (frictionless state) and nonzero values of $\mu$ (frictional states), respectively.
 Different functional forms for shear stress and particle pressure (note the leading $\phi^2$ term in \eqref{eq:P_phi_mu})
  leads to a stress ratio $\mu_{\rm bulk} = \sigma/\Pi$ (or $q = \Pi/\sigma$ in suspension flow modeling~\citep{Morris_1999}) being volume fraction dependent,
 consistent with the experimental results of Boyer {\it et al.} \citep{Boyer_2011}. 

%
%
%
  The friction-dependent parameters are found empirically from our simulations to be expressible as functions of $\mu$:  
     \begin{subequations}\label{phi_alpha_beta_K2_mu}
    \begin{equation}\label{phij_mu}
    \phi_{\rm J} (\mu) = \phi_{\rm J}^0 -(\phi_{\rm J}^0 - \phi_{\rm J}^\infty ) \exp(-\mu_\phi/\mu)~,
   \end{equation}
  \begin{equation}\label{alpha_mu}
      \alpha (\mu) = \alpha^0 +(  \alpha^\infty -\alpha^0 ) \exp(-\mu_\alpha/\mu)~,
   \end{equation}
       \begin{equation}\label{beta_mu}
  \beta (\mu) = \beta^0 +(  \beta^\infty -\beta^0 ) \exp(-\mu_\alpha/\mu)~,
   \end{equation}
  \begin{equation}\label{K2_mu}
      K_2 (\mu) = K_2^0 +(  K_2^\infty -K_2^0 ) \exp(-\mu_\alpha/\mu)
   \end{equation}
   \end{subequations}
as shown by the fits in Fig.~\ref{fig:consts_fric}.  The  fitting parameters $\mu_\phi$ and $\mu_\alpha$ are reported in table~\ref{table:param}; 
note that this table contains all friction-independent parameters of the model.  

   Next, we specify the flow curves utilizing a stress-dependent jamming volume fraction, 
   using an expression similar to that proposed by Wyart and Cates \citep{Wyart_2014}   
      \begin{subequations}\label{phi_alpha_beta_K2_str_mu}
   \begin{equation}\label{phi_str}
  \phi_{\rm m}(\tilde{\sigma},\mu) = \phi_{\rm J}(\mu)  f(\tilde{\sigma}) + \phi_{\rm J}^0 [ 1-f(\tilde{\sigma}) ]~,
   \end{equation}
   where $f(\tilde{\sigma})$ denotes the fraction of close particle interactions in which shear forces have overcome the stabilization
   repulsive force $F_0$ to achieve contact.
   We propose stress-dependent coefficients
  \begin{equation}\label{alpha_str}
    \alpha_{\rm m}(\tilde{\sigma},\mu) =  \alpha(\mu) f(\tilde{\sigma}) + {\alpha^0} (1-f(\tilde{\sigma}))~,
   \end{equation}
  \begin{equation}\label{beta_str_mu}
    \beta_{\rm m}(\tilde{\sigma},\mu) =  \beta(\mu)f(\tilde{\sigma}) + {\beta^0} (1-f(\tilde{\sigma}))~.
   \end{equation}
   \begin{equation}\label{K2_str_mu}
  K_{\rm m}(\tilde{\sigma},\mu) = K_2(\mu)  f(\tilde{\sigma} ) + K_2^0 \left (1-f(\tilde{\sigma}) \right )~.
   \end{equation}
   \end{subequations}

   Finally, using~\eqref{eq:eta_P_N2_phi}, \eqref{phi_alpha_beta_K2_mu}, 
    and \eqref{phi_alpha_beta_K2_str_mu} we propose the dependences of the rheological functions on $\tilde{\sigma},\phi$, and $\mu$:

     \begin{subequations}\label{eq:eta_P_N2_phi_str} 
       \begin{equation}\label{eq:eta_str_phi_mu}
     \eta_{\rm r} (\phi,\tilde{\sigma},\mu) = \alpha_{\rm m}(\tilde{\sigma},\mu) [\phi_{\rm m} (\tilde{\sigma},\mu) - \phi]^{-2} ~,
   \end{equation}
       \begin{equation}\label{eq:P_str_phi_mu}
     \frac{\Pi}{\eta_0\dot{\gamma}} (\phi,\tilde{\sigma},\mu) = \beta_{\rm m}(\tilde{\sigma},\mu) {\phi}^2  [\phi_{\rm m} (\tilde{\sigma},\mu) - \phi]^{-2} ~,
   \end{equation}
       \begin{equation}\label{eq:N2_str_phi_mu}
     \frac{N_2}{\eta_0\dot{\gamma}} (\phi,\tilde{\sigma},\mu) = K_{\rm m}(\tilde{\sigma},\mu) {\phi}^2  [\phi_{\rm m} (\tilde{\sigma},\mu) - \phi]^{-2} ~.
   \end{equation}
   \end{subequations}
 %

%
The divergences of the rheological functions described above -- viscosity, $N_2$, and $\Pi$ -- have the same algebraic sign at low and high stress. 
By contrast, $N_1$ presents a special case, in that it appears to have different signs under conditions dominated by 
lubrication and friction~\cite{Mewis_2011,Phung_1996,Cates_1998a,Foss_2000}.
We model $N_1$ as 
\begin{subequations}
\begin{equation}
\frac{N_1^{\rm L}}{\eta_0\dot{\gamma}}(\phi) = -K_1^{0} \phi^2 (\phi_{\rm J}^0-\phi)^{-2}
\end{equation}
\begin{equation}
\frac{N_1^{\rm H}}{\eta_0\dot{\gamma}}(\phi) = K_1(\mu) \phi^2 (\phi_{\rm J}(\mu)-\phi)^{-2}
\end{equation}
\end{subequations}
Now the stress-and volume fraction-dependent $N_1$ can be written as 
  \begin{subequations}\label{N1_phi_str_mu}
    \begin{equation}\label{eq:N1_str}
   \frac{N_1}{\eta_0\dot{\gamma}} (\tilde{\sigma},\phi) = K_1^{\rm m}(\tilde{\sigma}){\phi}^2 [\phi_{\rm m} (\tilde{\sigma}) - \phi]^{-2}~,
   \end{equation}
   where $K_1^{\rm m}(\tilde{\sigma})$ is given by
  \begin{equation}\label{K1_str}
   K_1^{\rm m}(\tilde{\sigma}) =  K_1(\mu) f(\tilde{\sigma}) - K_1^0 (1-f(\tilde{\sigma}))~.
   \end{equation}
     \end{subequations}
 The transition between the lubricated and frictionally dominated stress states is captured by the fraction of frictional interactions, $f(\tilde{\sigma})$, which we model as 
  $f(\tilde{\sigma}) = \exp \left[  -\tilde{\sigma}^{\ast}/\tilde{\sigma}\right]$, with $\sigma^{\ast}=1.45\sigma_0$ based on simulations here and previously published results \citep{Mari_2014,Guy_2015, Ness2016, Royer_2016, Hermes_2016}.    We assume that $f(\tilde{\sigma})$ does not depend on $\mu$.

   \section{Results}
\begin{figure*}
\centering
\subfigure[]{
\includegraphics[width=.45\textwidth]{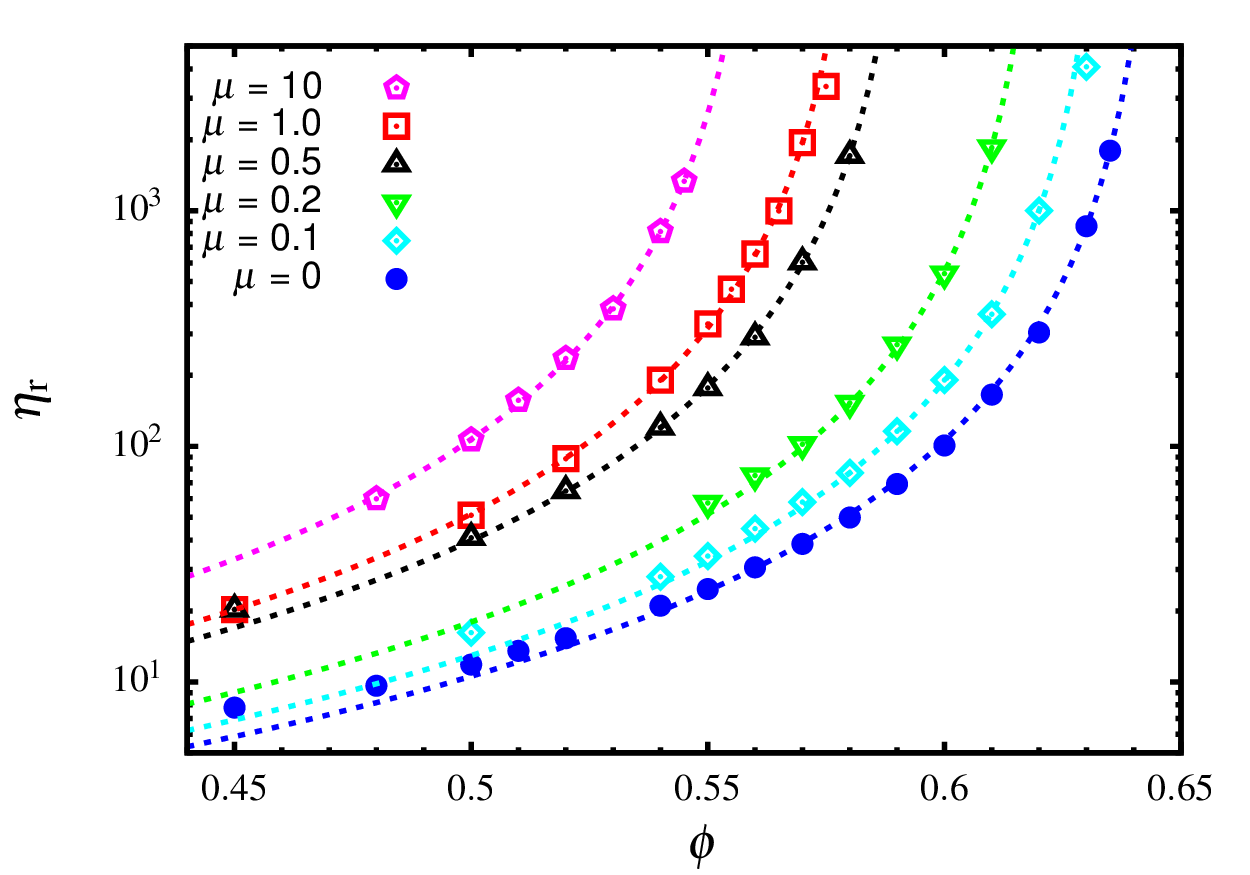}}
\subfigure[]{
\includegraphics[width=.45\textwidth]{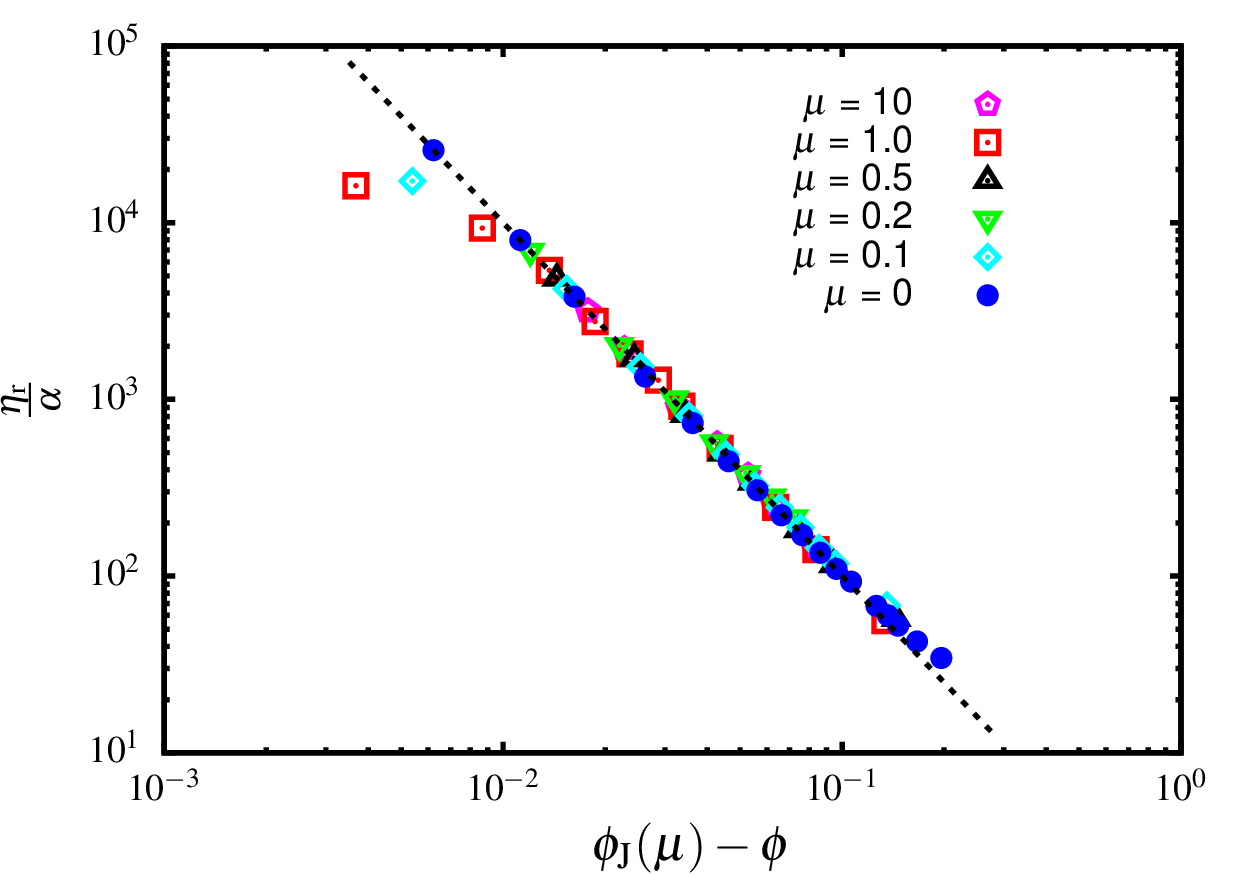}}
\subfigure[]{
\includegraphics[width=.45\textwidth]{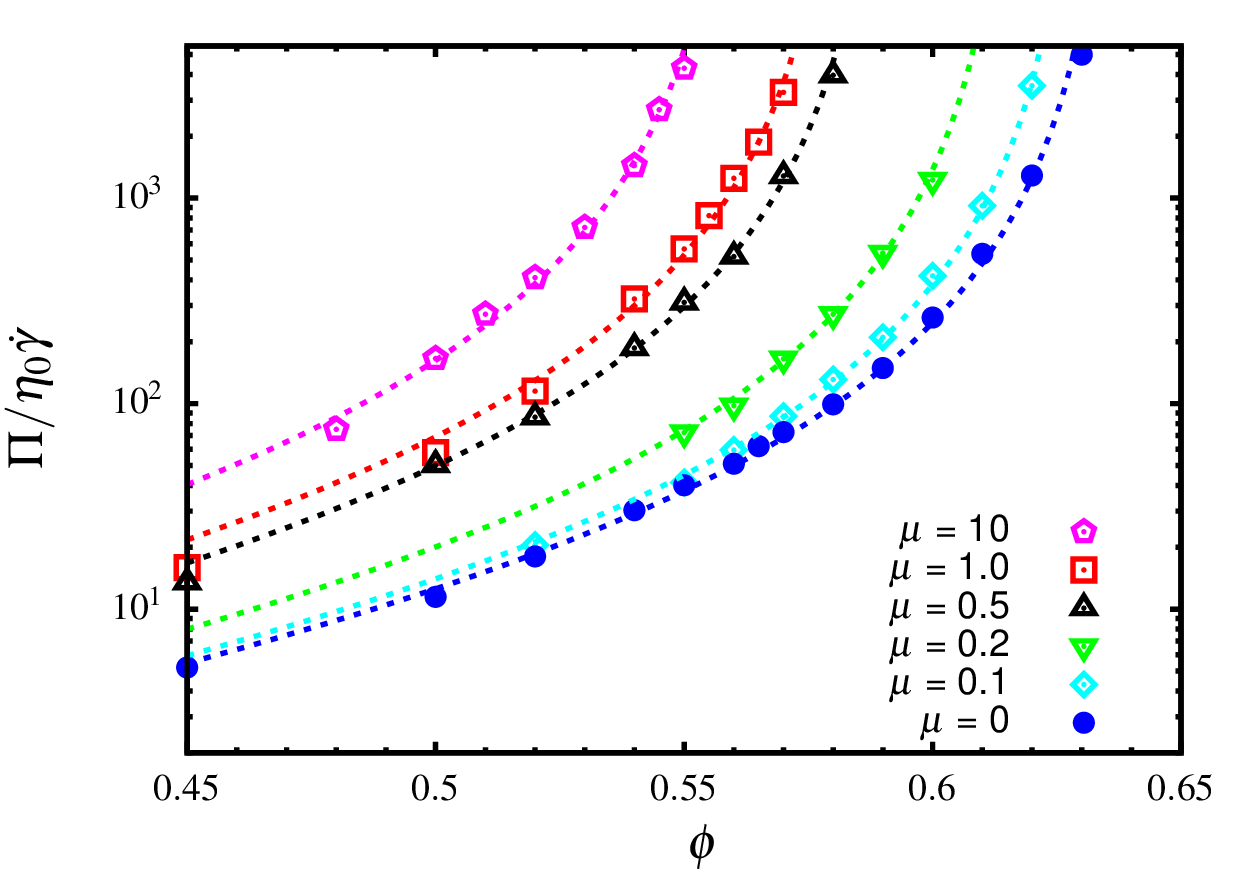}}
\subfigure[]{
\includegraphics[width=.45\textwidth]{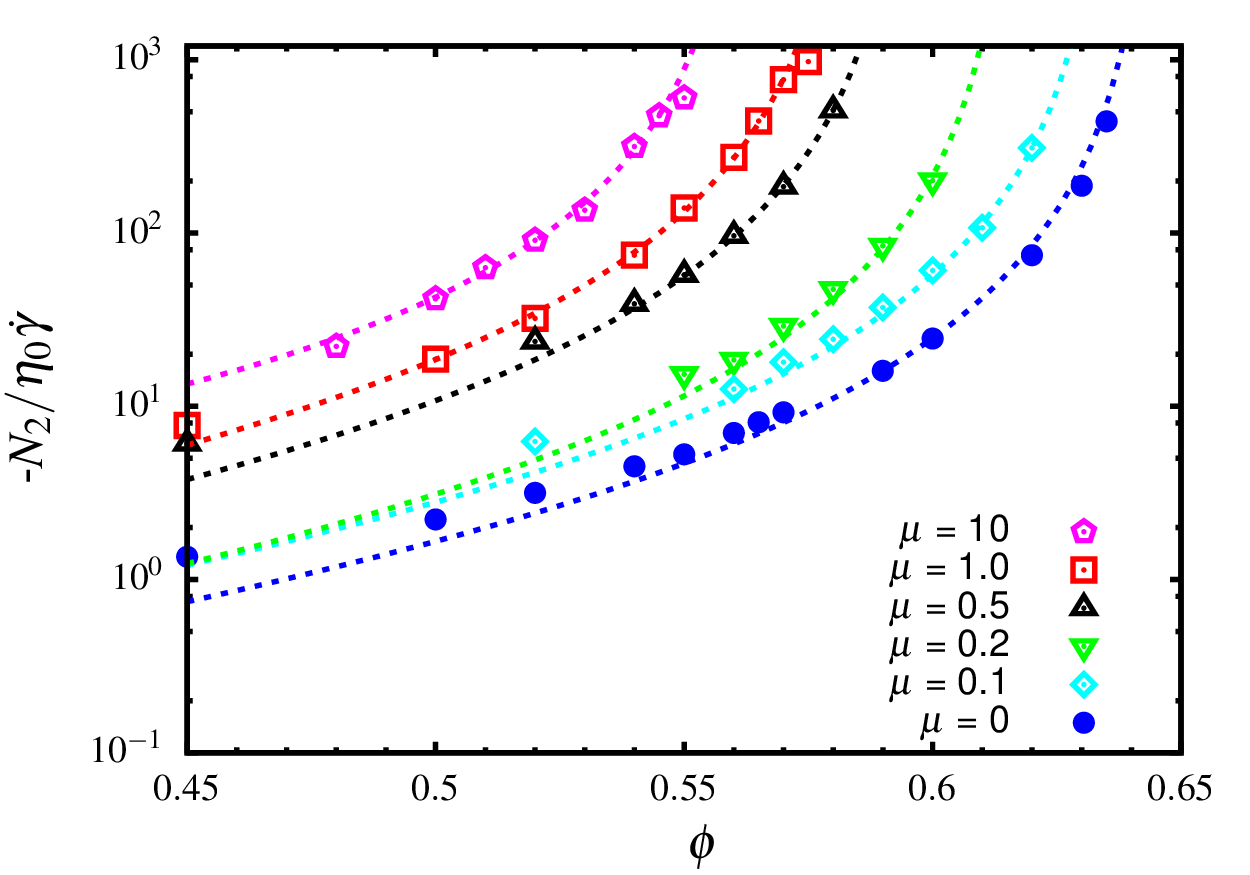}}
\caption{ 
(a) $\eta_r(\phi)$, (c) $\Pi/\eta_0\dot{\gamma}(\phi)$ and (d) $N_2/\eta_0\dot{\gamma}(\phi)$
for the rate-independent frictionless (low stress $0.1<\tilde{\sigma}<0.3$) and frictional (high stress $\tilde{\sigma}>10$) states from simulations, with different values of $\mu$ in the frictional case.
Filled symbols represent the frictionless state, while open symbols represent different interparticle friction coefficients.
Dashed lines in (a), (c) and (d) are fit to~\eqref{eq:eta_P_N2_phi}.
(b) Logarithmic plot of $\frac{\eta_{\rm r}}{\alpha}$ versus $(\phi_{\rm J}(\mu)-\phi)$ for different values of $\mu$. The dashed line is a guide to eye and shows power law -2.
}
\label{fig:fric_full}
\end{figure*}

Before turning to stress dependence, we show in Fig.~\ref{fig:fric_full} our simulation results for the stress-independent $\eta_{\rm r}$, $\Pi$ and $N_2$
for rate-independent states. 
These agree well with~\eqref{eq:eta_P_N2_phi} for all values of $\mu$. 
The viscosity at large $\phi$ is well represented by $(\phi_{\rm J}-\phi)^{-2}$ independent of $\mu$, as shown in Fig.~\ref{fig:fric_full}b;
$\phi_{\rm J}$ is obtained
by a least-squares fit of~\eqref{eq:etaL_phi} and~\eqref{eq:etaH_phi} to the volume fraction dependence of the viscosity at low 
$(0.1<\tilde{\sigma}<0.3)$ and high $(\tilde{\sigma}>10)$
stresses, respectively.

\begin{figure*}
\centering
\subfigure[]{
\includegraphics[width=.45\textwidth]{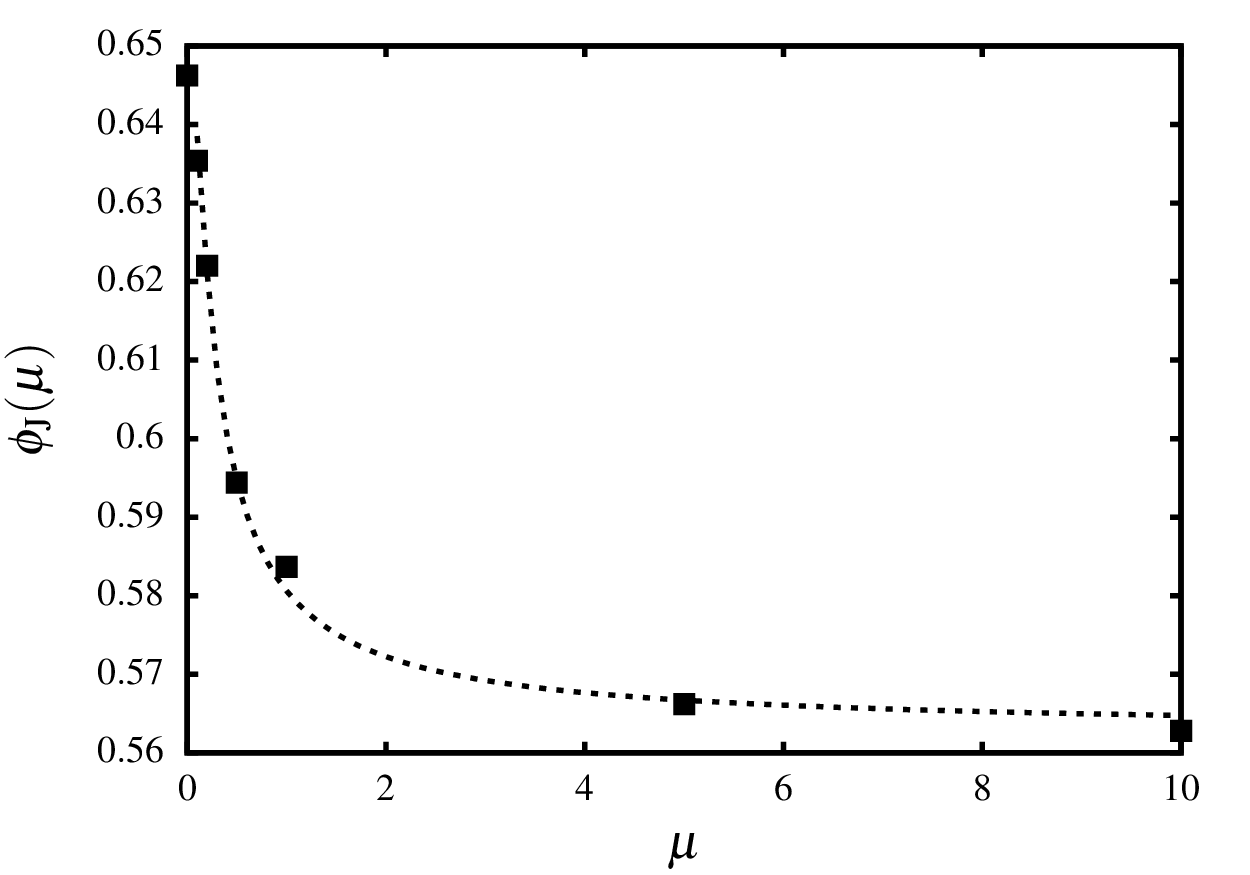}}
\subfigure[]{
\includegraphics[width=.45\textwidth]{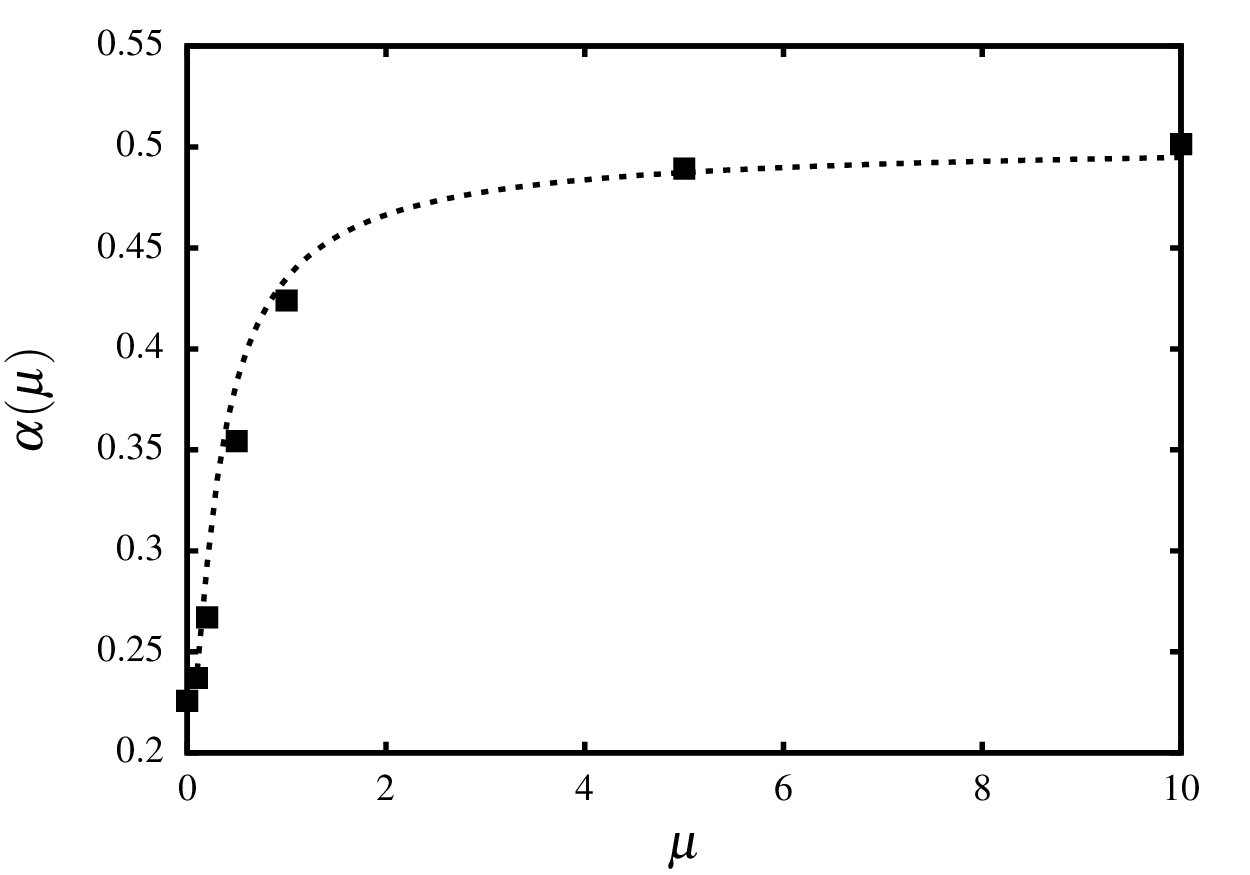}}
\subfigure[]{
\includegraphics[width=.45\textwidth]{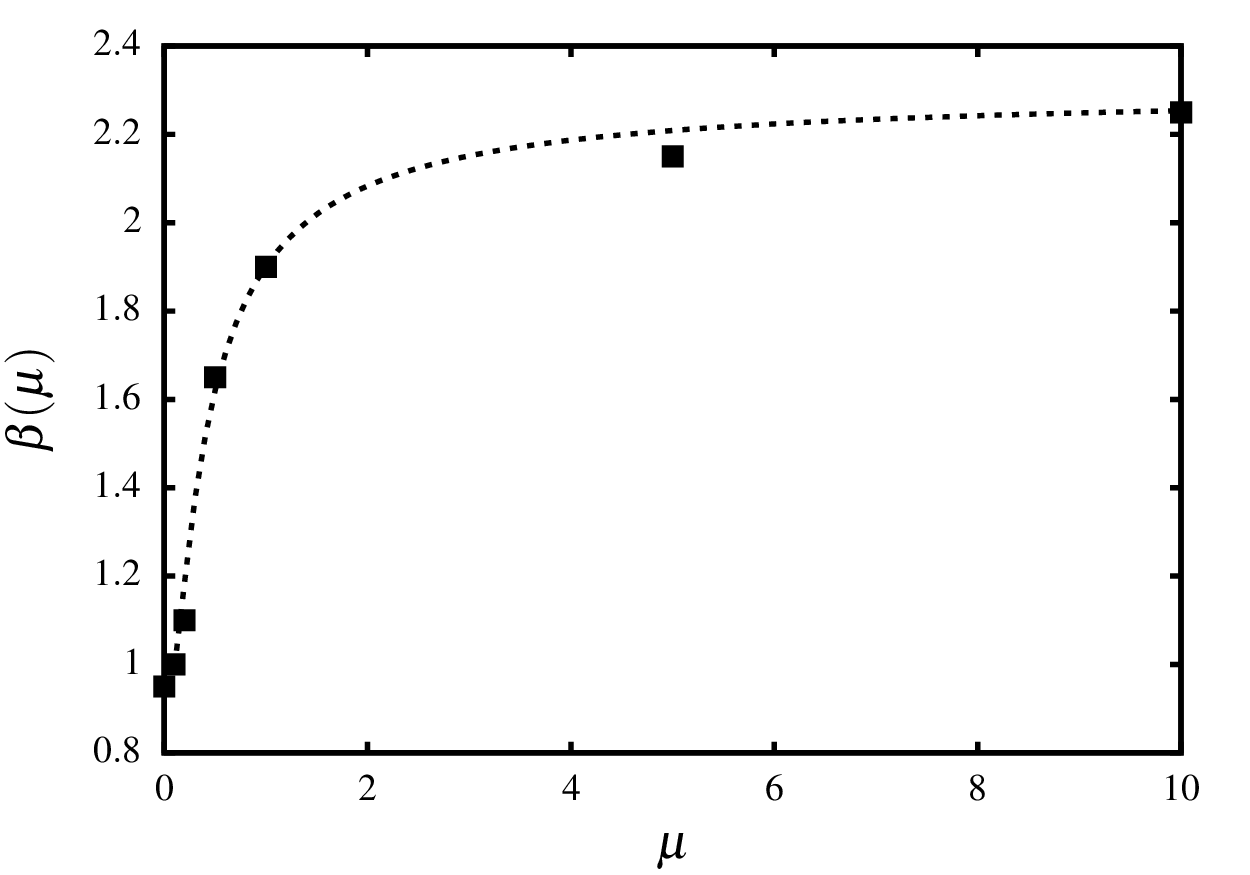}}
\subfigure[]{
\includegraphics[width=.45\textwidth]{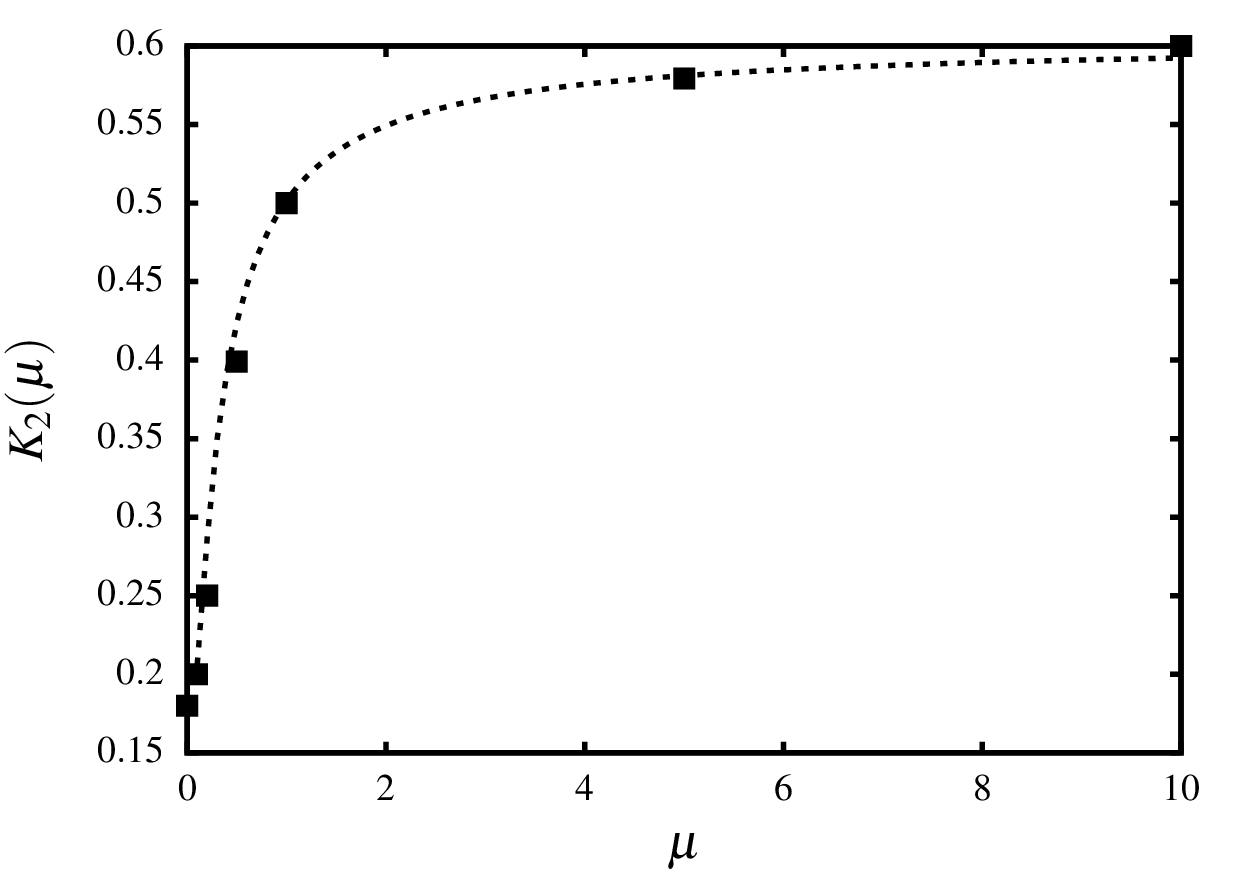}}
\caption{ 
(a) Jamming volume fraction $\phi_{\rm J}(\mu)$ (square) plotted as a function of $\mu$.
Other friction dependent constants (b) $\alpha(\mu)$, (c)  $\beta(\mu)$ and (d) $K_2(\mu)$ plotted as a function of $\mu$.
Dashed lines in (a), (b), (c), and (d) are best fits of the data to~\eqref{phij_mu}, \eqref{alpha_mu}, \eqref{beta_mu}, and \eqref{K2_mu} respectively.
}
\label{fig:consts_fric}
\end{figure*}

The friction-dependent constants are found to fit well to exponential functions of the form proposed
in~\eqref{phi_alpha_beta_K2_mu} as shown in 
Fig.~\ref{fig:consts_fric}; values of these parameters are presented in the Table~\ref{table:param}.
Because $N_1$ proves more difficult to reliably simulate
(or experimentally measure~\cite{Lee_2006a,Lootens_2005,Cwalina_2014}), we defer its consideration to a later section.

\begin{table*}
 \caption{$\mu$ independent model constants}
\begin{center}
 \begin{tabular}{|c c c c c c c c c c c|} 
 \hline
 $\phi_{\rm J}^0$ & $\phi_{\rm J}^\infty$ & $\alpha^0$ & $\alpha^\infty$  & $\beta^0$ & $\beta^\infty$    & $K_2^0$ & $K_2^\infty$   & $\mu_\phi$ & $\mu_\alpha$ & $K_1^0$\\ [0.5ex] 
 \hline
 0.646 & 0.562 & 0.225 & 0.510 & 0.95 & 2.25 &   0.18 & 0.61 & 0.24 &0.275 &0.055 \\  [1ex] 
 \hline
\end{tabular}
\label{table:param}
\end{center}
\end{table*}

  \subsection{Rate dependent viscosity} 
\begin{figure*}
\centering
\subfigure[]{
\includegraphics[width=.32\textwidth]{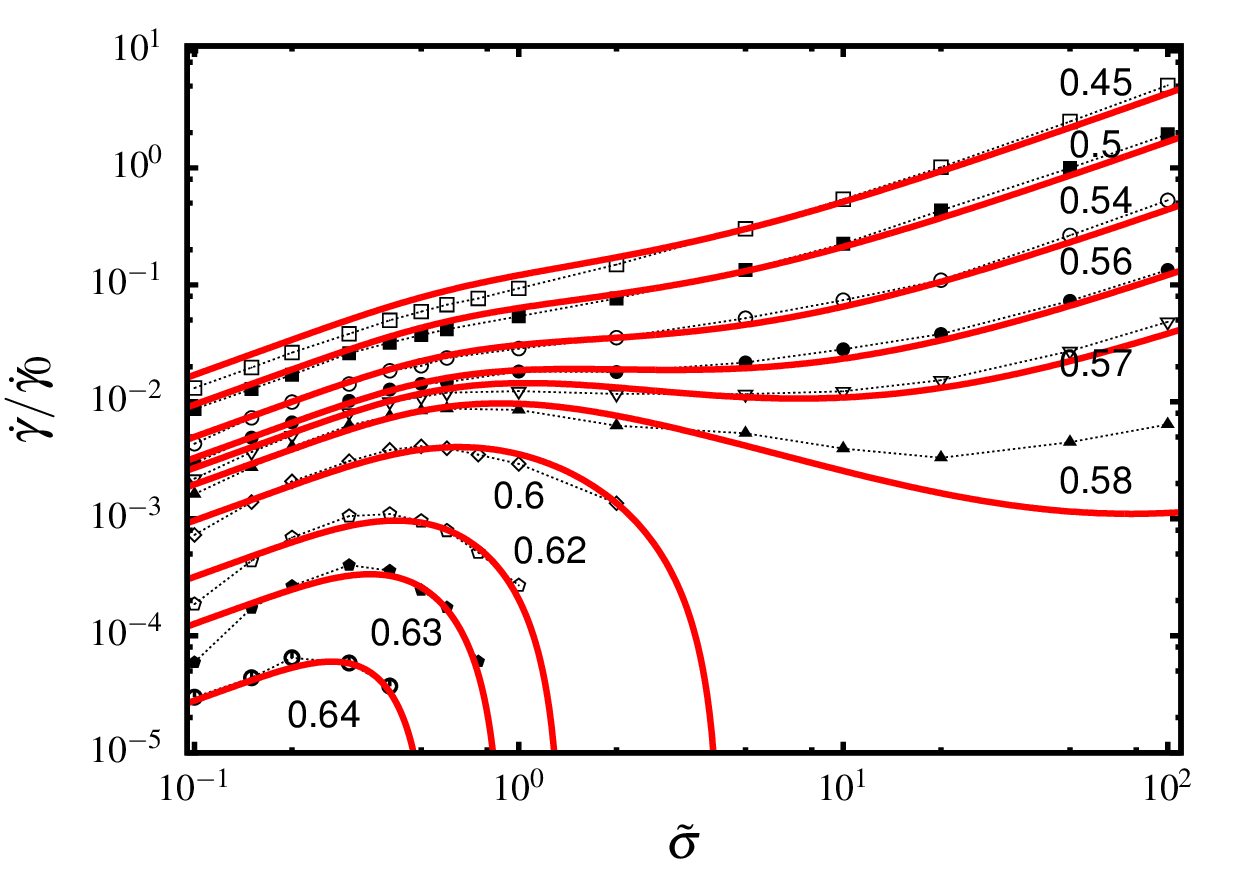}}
\subfigure[]{
\includegraphics[width=.32\textwidth]{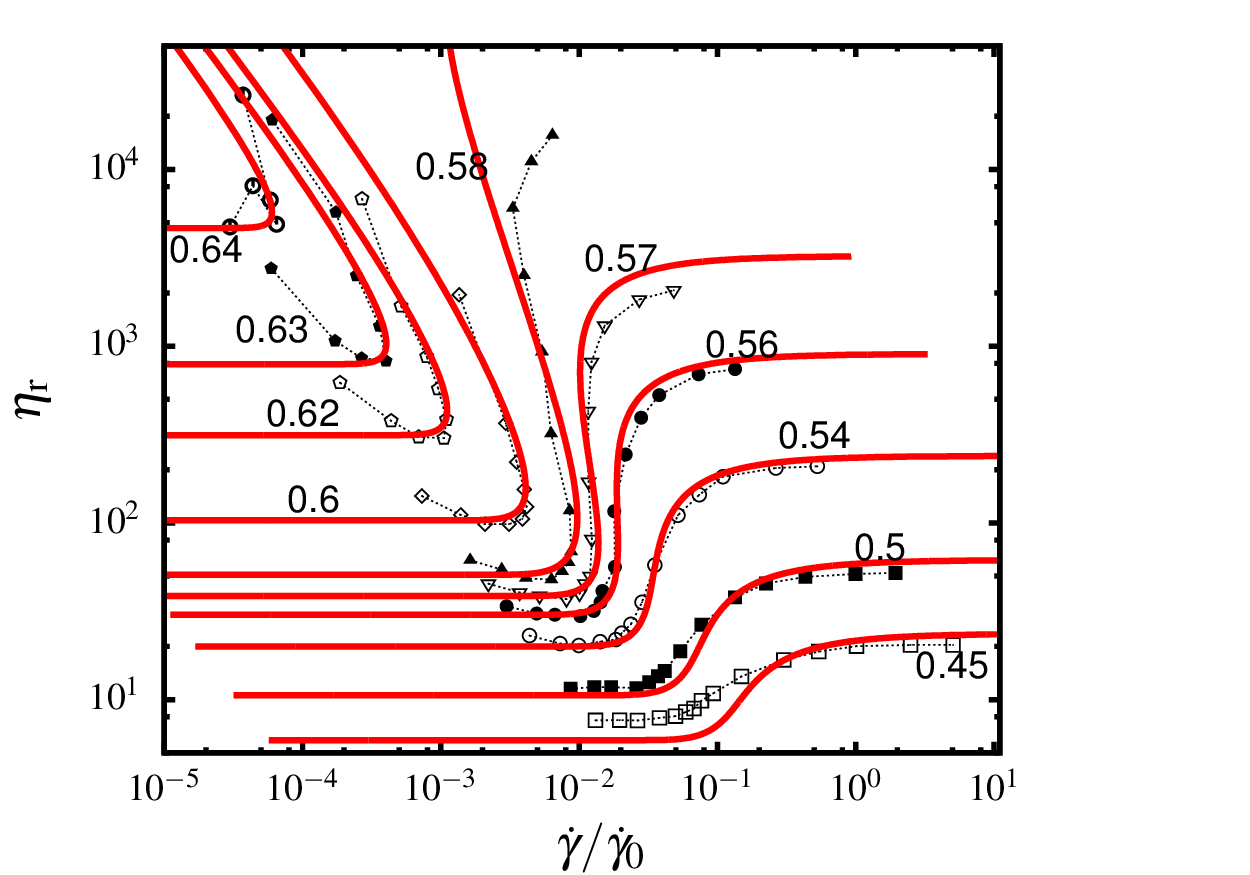}}
\subfigure[]{
\includegraphics[width=.32\textwidth]{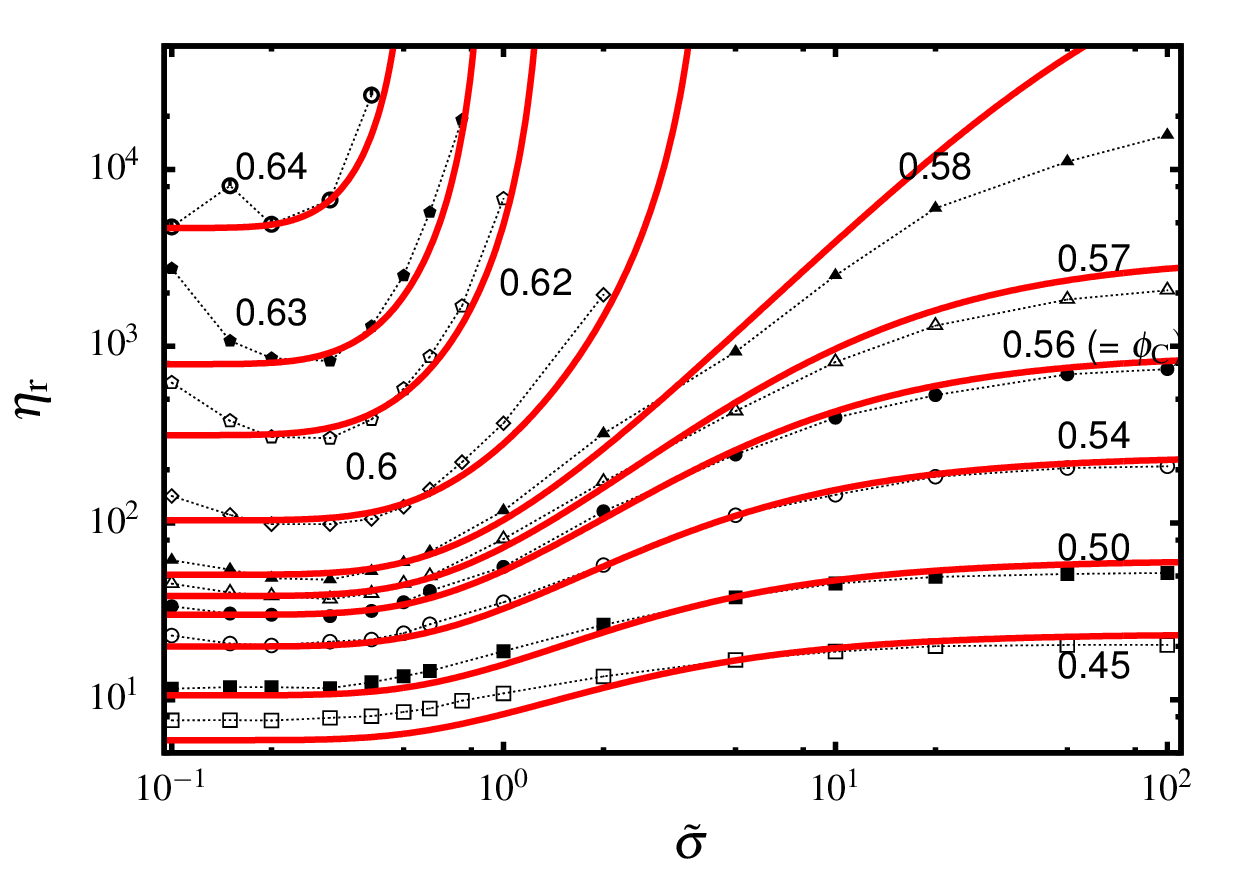}}
\caption{ Steady state flow curves for several values of volume fraction $\phi$ at $\mu=1$.
(a) The dimensionless  rate $\dot{\gamma}/\dot{\gamma}_0$ as a function of dimensionless applied stress $\tilde{\sigma}$.
Continuous shear thickening (CST) observed at low volume fraction ($\phi<0.56$) is associated with monotonic flow curves,
discontinuous shear thickening (DST) appears as non-monotonic flow curves for $0.56\le \phi \le 0.58$, and for $\phi \ge 0.59$ the system is shear jammed at high stress.
(b) The same data plotted as $\eta_{\rm r} (\dot{\gamma}/\dot{\gamma}_0)$ flow curve.
(c) Relative viscosity $\eta_{\rm r}$ as a function of dimensionless applied stress $\tilde{\sigma}$.
The symbols are simulation data with dashed lines provided to guide the eye. The solid lines are predictions from~\eqref{eq:eta_str_phi_mu}.
}
\label{fig:visc_mu1p0}
\end{figure*}
%
To develop a sense of the entire flow behavior, we present in Fig.~\ref{fig:visc_mu1p0} the viscosity data with interparticle friction coefficient $\mu = 1$. This value of $\mu$ is comparable to the experimentally measured values of 
 Fernandez {\it et al.} \cite{Fernandez_2013}, where $\mu$ is in the range $0.6$--$1.1$ for polymer brush-coated quartz particles of diameter
 $2a \sim10$ $\mu \rm m$, but is higher than the value of 0.5 reported by Comtet {\it et al.}~\cite{Comtet_2017}.
 The data are presented in the forms $\dot{\gamma}(\tilde{\sigma})/\dot{\gamma}_0$, $\eta_{\rm r}(\dot{\gamma}/\dot{\gamma}_0)$,  and $\eta_{\rm r}(\tilde{\sigma})$ for a range of values of $\phi \ge 0.45$. 
 Figure~\ref{fig:visc_mu1p0}a shows that 
 for volume fractions $\phi<0.56$ the $\dot{\gamma}(\tilde{\sigma})/\dot{\gamma}_0$ curves are monotonic and show continuous shear thickening for $0.3 \le \tilde{\sigma} \le 10$.
   At $\phi=0.56$, the curve exhibits the first sign of non-monotonicity: we define $\phi_{\rm C} \doteq 0.56$ to signify the DST onset volume fraction.
   For $\phi=0.57$ and  $\phi = 0.58$, the slope is negative (i.e., $d \dot{\gamma}/ d \tilde{\sigma}<0$) for intermediate stress but crosses over to a positive slope for $\tilde{\sigma}>10$, corresponding to the S-shaped $\eta_{\rm r}(\dot{\gamma}/\dot{\gamma}_0)$ curves shown in Fig.~\ref{fig:visc_mu1p0}b.
 The viscosity as modeled by~\eqref{eq:eta_str_phi_mu} is shown by solid lines and agrees with the simulation data except at high stresses at $\phi=0.58$. This discrepancy is due to the closeness
 to $\phi_{\rm J}(\mu)$ given by~\eqref{phij_mu}, which slightly underestimates the jamming volume fraction for $\mu=1$.

 Although plotted as a function of shear rate, it is important to note that these simulations were performed at fixed shear stress:
   DST would be observed at fixed rate for this range of volume fraction ($0.56 \le \phi \le 0.58$)
   as S-shaped curves are not accessible in a rate-controlled scenario \citep{Wyart_2014,Mari_2015}.
   For this range of $\phi$, where DST is observed between two {\it flowing} states,
   we term the discontinuous shear thickening as \textquotedblleft pure DST\textquotedblright.
   However for $\phi \ge 0.59$, the upper, i.e. high stress, branch of the S-shaped $\dot{\gamma}(\tilde{\sigma})$ curves (Fig.~\ref{fig:visc_mu1p0}a) is not accessible, signifying that the 
   suspension enters a shear jammed (SJ) state above $\tilde{\sigma}_{\rm J}(\phi)$.
     The suspension is flowable for low stress, but is jammed for $\tilde{\sigma}> \tilde{\sigma}_{\rm J}(\phi)$;
     $\tilde{\sigma}_{\rm J}(\phi)$ decreases with increasing $\phi$.
     We term such discontinuous shear thickening, in which the thickening continues until reaching a shear-jammed state, as \textquotedblleft DST-SJ\textquotedblright.
     When the traditional flow curve $\eta_{\rm r}(\dot{\gamma}/\dot{\gamma}_0)$ is plotted for the same data in Fig.~\ref{fig:visc_mu1p0}b, we observe non-monotonicity for $\phi \ge 0.56$.
    This data, when presented in the form $\eta_{\rm r}(\tilde{\sigma})$, shows that the onset stress for shear thickening $\tilde{\sigma}_{\rm ST} \approx 0.3$ is roughly 
    independent of volume fraction, as observed in previous studies \cite{Maranzano_2001,Larsen_2010,Brown_2014,Mari_2014}.
    To characterize shear thickening as CST or DST, we fit $\eta_{\rm r}(\tilde{\sigma})$ in the thickening regime to $\eta_{\rm r} \sim \tilde{\sigma}^\zeta$:
     $\zeta<1$ implies continuous shear thickening, while $\zeta = 1$ implies the onset of DST, and larger $\zeta$ are in the DST region.
    We also observe some shear thinning at low stress as a result of the short-range electrostatic repulsion \citep{Mari_2014}.

\begin{figure*}
\centering
\subfigure[]{
\includegraphics[width=.32\textwidth]{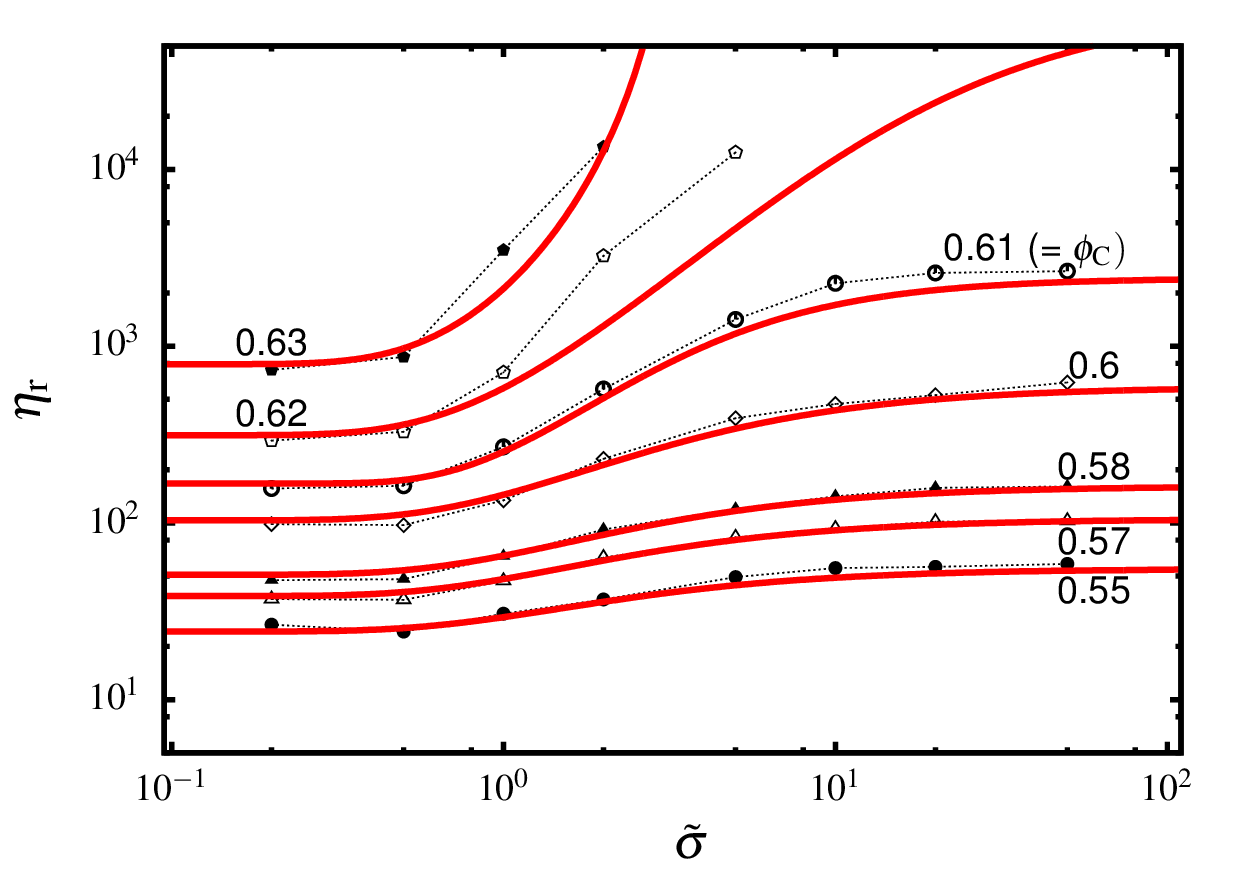}}
\subfigure[]{
\includegraphics[width=.32\textwidth]{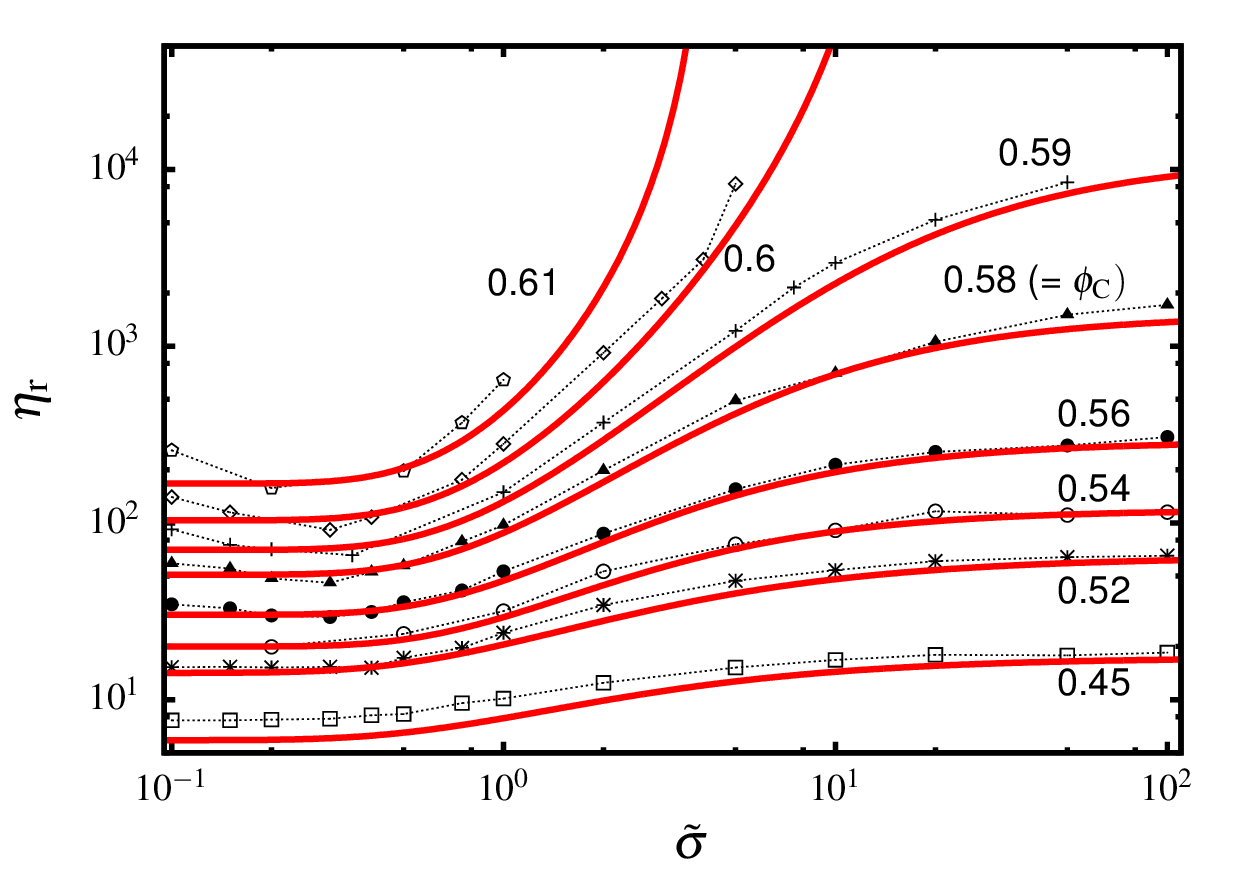}}
\subfigure[]{
\includegraphics[width=.32\textwidth]{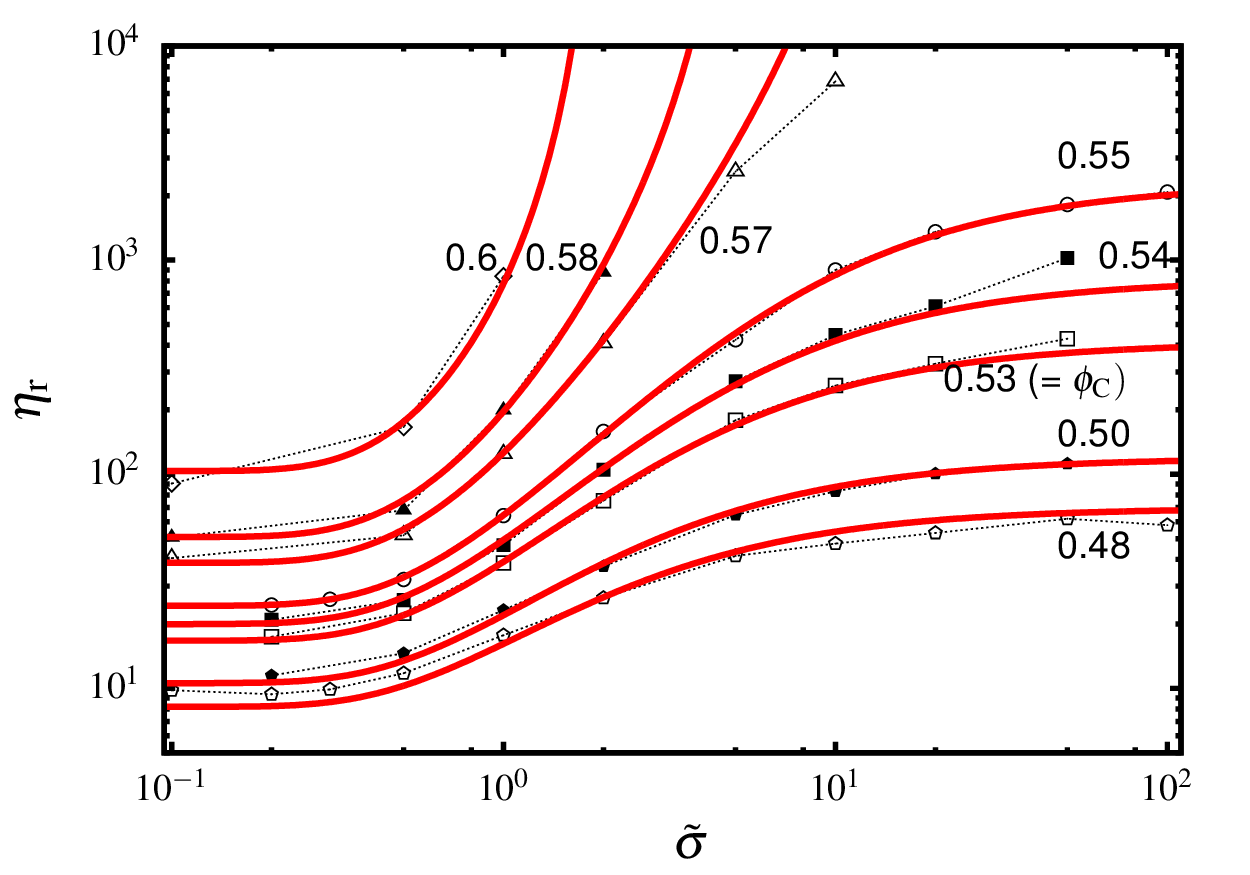}}
\caption{ Varying friction coefficient: steady state relative viscosity $\eta_{\rm r}$ plotted against scaled applied stress $\tilde{\sigma} = \sigma/\sigma_0$  for friction coefficient $\mu = $ (a) 0.2, (b) 0.5 and (c) 10 for several values of volume fractions as mentioned.
$\phi_{\rm C}$ denotes the volume fraction for the onset of DST.
Symbols and dashed lines indicate the simulation data while solid lines are predictions from~\eqref{eq:eta_str_phi_mu}.
}
\label{fig:Visc_fric_full_comp}
\end{figure*}

    The viscosity obtained from~\eqref{eq:eta_P_N2_phi_str} is shown in Fig.~\ref{fig:Visc_fric_full_comp} along with simulation data for different values of $\mu$.
 The model is in excellent agreement with the data. The phenomenology is the same for any value of $\mu$ expect for $\mu=0$.
 With increasing $\mu$, the DST onset volume fraction $\phi_{\rm C}$ decreases.   
 
\subsection{Flow state diagram}   
The shear rheology described above is controlled by three dimensionless parameters, namely the solid volume fraction $\phi$, 
dimensionless shear stress $\tilde{\sigma}$, and interparticle friction coefficient $\mu$.

\begin{figure*}
\centering
\subfigure[]{
\includegraphics[width=.48\textwidth]{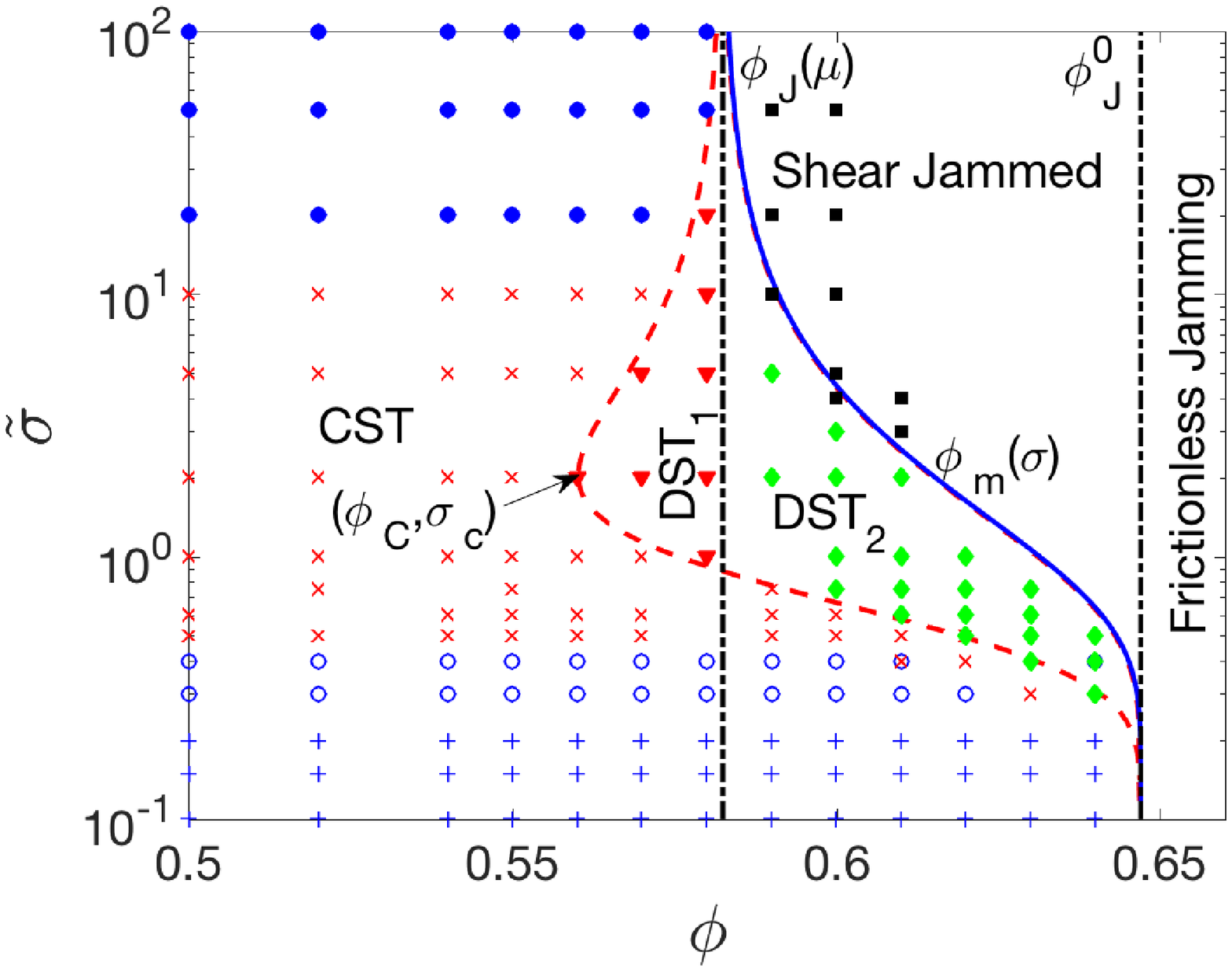}}
\subfigure[]{
\includegraphics[width=.48\textwidth]{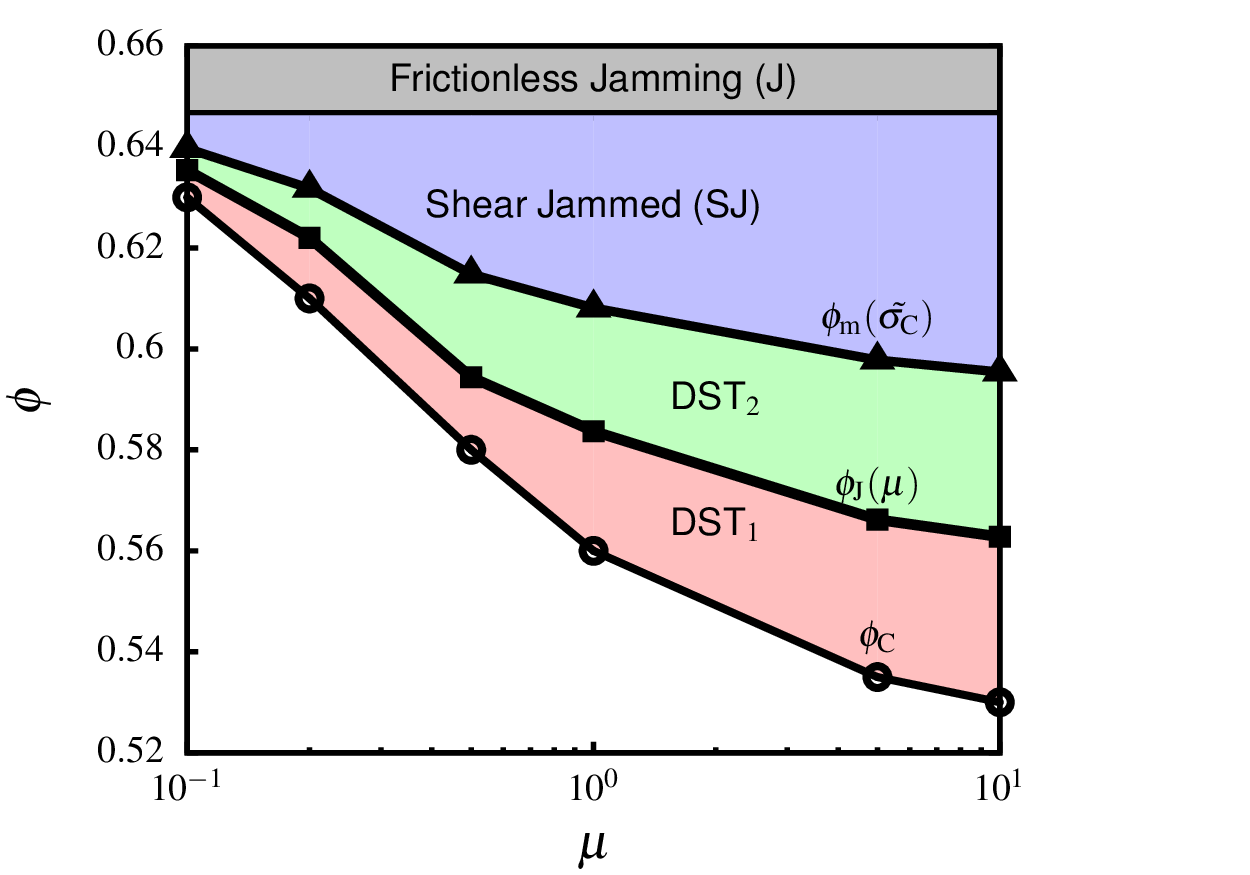}}
\caption{
 Flow state diagram $\tilde{\sigma}, \phi, \mu$ shown in two different projections:
  (a) $\phi$ - $\tilde{\sigma}$ plane for a constant interparticle friction $\mu=1.0$. 
  The blue solid line is the stress dependent jamming line $\phi_{\rm m} (\tilde{\sigma})$, while the dashed red line is the DST line and shows locus of points where $\frac{d{\dot{\gamma}}}{d\sigma} = 0$.
   Dot-dashed black lines represent $\phi_{\rm J}({\mu})$ and $ \phi_{\rm J}^0$. 
   Symbols represent different states of the suspension: shear thinning (blue plus), rate-independent (blue circles), continuous shear thickening (red crosses), pure DST (red diamonds), DST-SJ (green diamonds) and shear jammed states
   (black squares). Along the $\phi$ axis, there are three special densities: $\phi_{\rm C}$ below which there is no DST region,
    $\phi_{\rm J}({\mu})$, below which there is no shear jamming, and $ \phi_{\rm J}^0$, above which isotropically jammed states exist.
    Corresponding to $\phi_{\rm C}$, DST exists for only one value of  stress $\tilde{\sigma}_{\rm C}$, while for $\phi>\phi_{\rm C}$ DST
    exists for a range of stress values.
(b) $\mu$ - $\phi$ plane for a constant stress $\tilde{\sigma}_{\rm C}$. Circles represent the DST onset volume fraction $\phi_{\rm C}$,
squares represent frictional jamming point $\phi_{\rm J}({\mu})$, triangles show $\phi_{\rm m}(\tilde{\sigma}_{\rm C})$.
In between $\phi_{\rm C}$
 and $\phi_{\rm J}({\mu})$ pure DST is observed, 
 while the green region between $\phi_{\rm J}({\mu})$ and $\phi_{\rm m}(\tilde{\sigma}_{\rm C})$ DST-SJ is observed.
 In the blue region above $\phi_{\rm m}(\tilde{\sigma}_{\rm C})$, the suspension is in a shear-jammed state.}
\label{fig:phase_diagram}
\end{figure*}

The results discussed are presented in a flow state diagram.  
As this depends on three variables, we present two views: 
 in the $\phi$ - $\tilde{\sigma}$ plane for $\mu=1$, and in the $\mu$ - $\phi$  plane for stress  $\tilde{\sigma}_{\rm C}$. 
 Figure~\ref{fig:phase_diagram}a displays the observed flow state diagram in the $\phi$ - $\tilde{\sigma}$ plane, and here we identify three volume fractions:
  $\phi_{\rm C}$, $\phi_{\rm J}(\mu)$, and $\phi_{\rm J}^0$. 
 Vertical lines represent frictional $\phi_{\rm J}(\mu)$ and frictionless $\phi_{\rm J}^0$ jamming points.
In the lower part of the diagram, where the stress is too low to overcome the interparticle repulsive force, friction between close particles is not activated
 and hence the rheology diverges at $\phi_{\rm J}^0$. 
 However, in the upper part of the  flow state diagram where the stress is large most of the close interactions (or \textquotedblleft contacts\textquotedblright) are frictional
 which leads to divergence of 
 viscosity and other rheological functions at $\phi_{\rm J}(\mu) < \phi_{\rm J}^0$.
 In the two extremes, the viscosity in the model is rate independent.
However, in the simulations at low stress, the finite range of repulsion leads to a larger apparent particle size, and the competition between short-range repulsion and external applied
   stress creates a shear thinning behavior, at the conditions indicated by the $+$ symbols.
For intermediate stress, continuous shear thickening is observed in the range of $\phi<\phi_{\rm C}$.
For $\phi_{\rm C} \le \phi < \phi_{\rm J}(\mu)$,   \textquotedblleft pure DST\textquotedblright 
is observed (shown by triangles).
In this range of $\phi$, the dashed line is the envelope of the pure DST states, with $(\phi_{\rm C}, \tilde{\sigma}_{\rm C})$ being the point with the minimum $\phi$
value along this line. This line is determined as the locus of points for which $d{\dot{\gamma}}/d{\tilde{\sigma}}=0$ in a flow curve $\dot{\gamma}(\tilde{\sigma})$
as shown in Fig.~\ref{fig:visc_mu1p0}a: there are two such points on a curve for any $\phi>\phi_{\rm C}$ and coalescence of these two points occurs at a  
 critical point ($\phi_{\rm C}, \tilde{\sigma}_{\rm C}$).
\par

For $\phi> \phi_{\rm J}(\mu)$, the upper boundary of DST states is the stress-dependent jamming line $\phi_{\rm m}(\tilde{\sigma})$.
The jamming line separates the DST regime (diamonds) from conditions yielding a solid-like shear-jammed state (squares). 
The distinction between two types of DST regimes is based on differences in the high stress state, which is flowable in the pure DST regime 
(for $\phi_{\rm C} \le \phi < \phi_{\rm J}(\mu)$) and jammed for DST-SJ (for $\phi_{\rm J}(\mu) \le \phi < \phi_{\rm J}^0$).
The minimum stress required to observe DST and shear-jammed (SJ) states decreases with increasing $\phi$, 
and eventually these curves converge and the minimum stress for jamming tends to zero as the frictionless jamming point $\phi_{\rm J}^0$ is approached. 
%
%

Figure~\ref{fig:phase_diagram}b displays the flow state diagram in the $\mu$ - $\phi$ plane for a constant stress $\tilde{\sigma}_{\rm C}$,
 the minimum stress for DST; we note that $\tilde{\sigma}_{\rm C}$ is roughly independent of $\mu$. 
Here, we see a demonstration of how the volume fractions $\phi_{\rm C}$, $\phi_{\rm J}(\mu)$, and $\phi_{\rm m}(\tilde{\sigma}_{\rm C})$ decrease as a function of $\mu$.
The region enclosed between $\phi_{\rm C}$ and $\phi_{\rm J}(\mu)$ broadens at larger $\mu$, illustrating that the range of $\phi$ over which  \textquotedblleft pure DST\textquotedblright  is observed broadens with increasing interparticle friction.
 For the range of volume fractions 
$\phi_{\rm J}(\mu) < \phi < \phi_{\rm m}(\tilde{\sigma}_{\rm C})$, the suspension is in the DST-SJ region, and above $\phi_{\rm m}(\tilde{\sigma}_{\rm C})$, the  
system is in shear-jammed state at the imposed stress.
%

  %
  \subsection{Rate dependent normal stresses}  
\begin{figure*}
\centering
\subfigure[]{
\includegraphics[width=.48\textwidth]{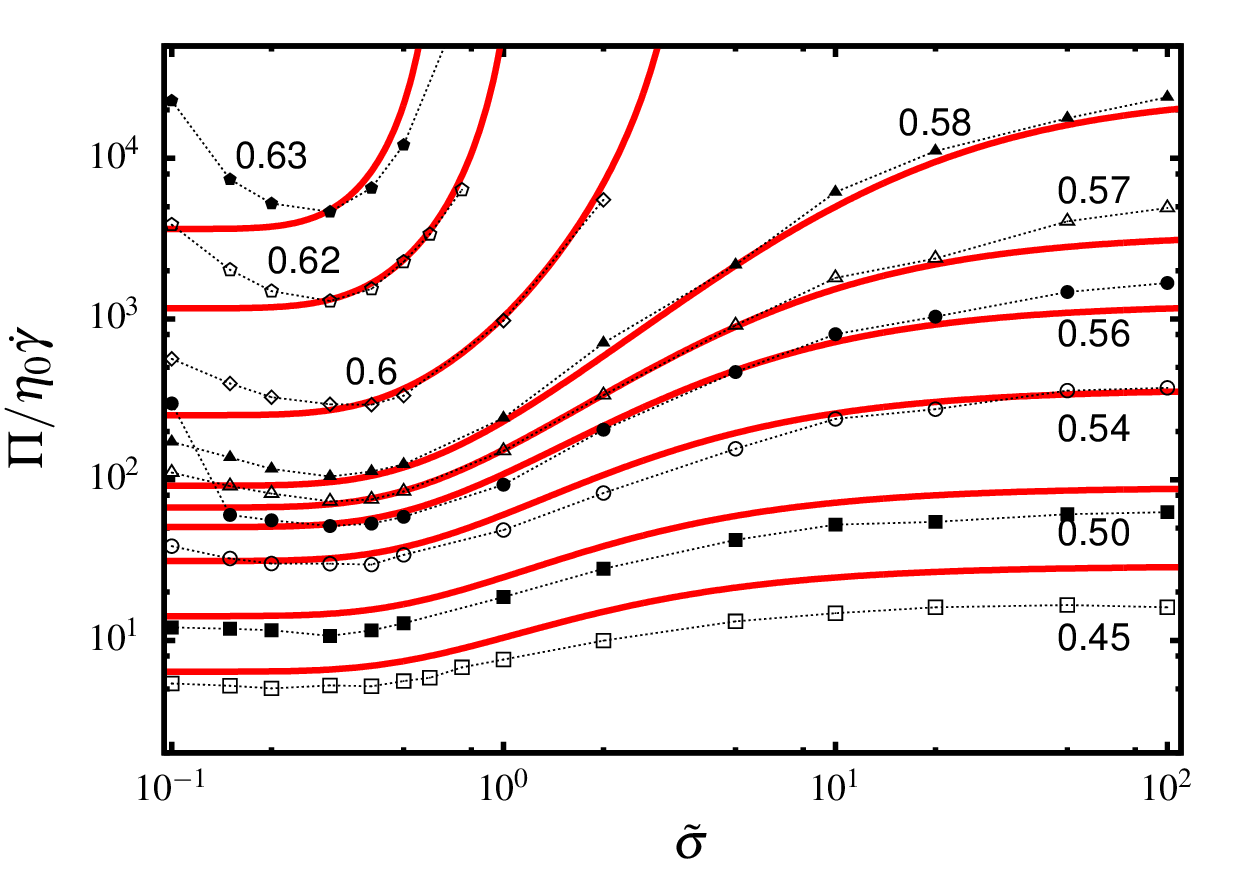}}
\subfigure[]{
\includegraphics[width=.48\textwidth]{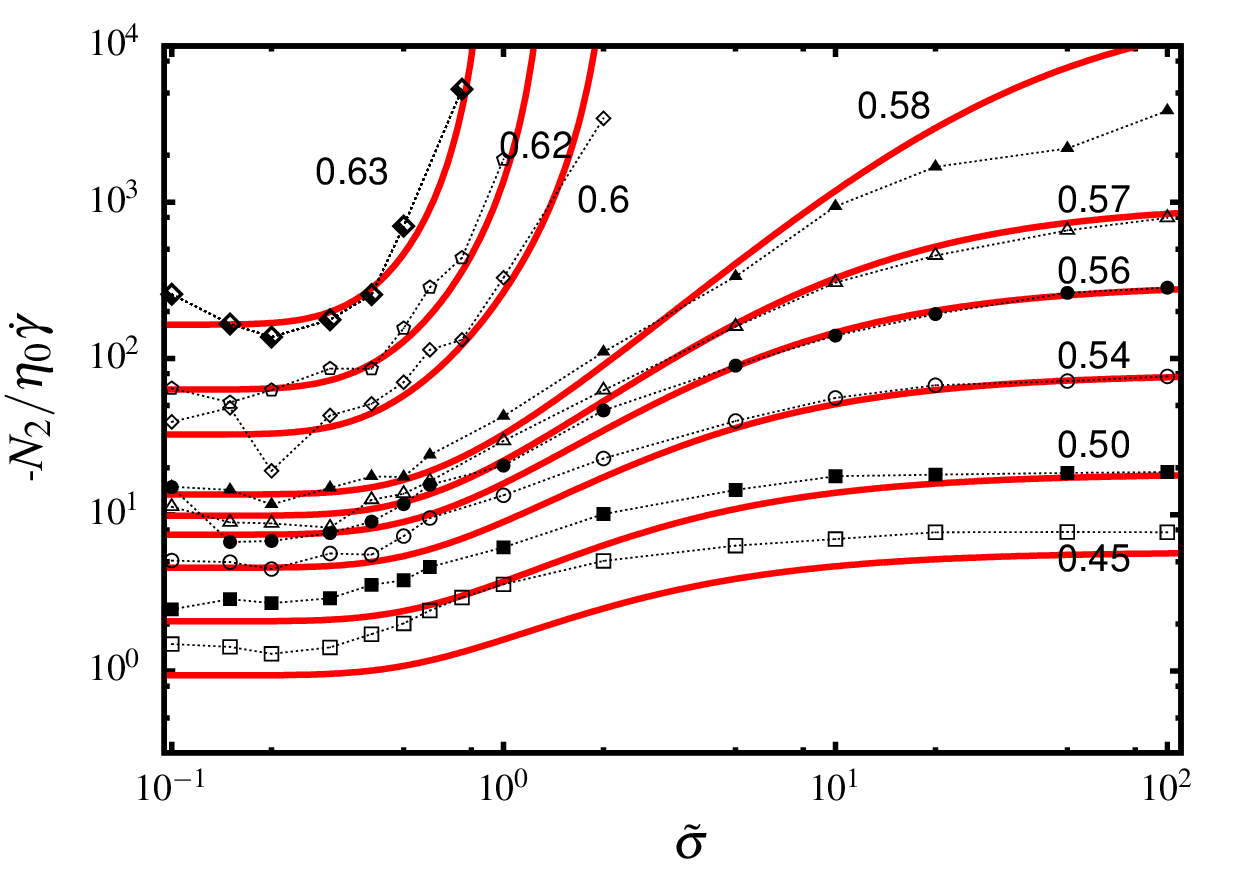}}
\caption{ Steady state (a) particle pressure $\Pi/\eta_0\dot{\gamma}$, and (b)
second normal stress difference $N_2/\eta_0\dot{\gamma}$ plotted against applied stress $\tilde{\sigma} = \sigma/\sigma_0$  for
$\mu = 1$.
Symbols and dashed lines indicate the simulation data while solid lines are predictions from~\eqref{eq:P_str_phi_mu} and \eqref{eq:N2_str_phi_mu}.
}
\label{fig:P_N2_mu1p0}
\end{figure*}

The simulation data along with the model predictions for the particle pressure $\Pi/\eta_0\dot{\gamma}$ and second normal stress difference $N_2/\eta_0\dot{\gamma}$, are presented in Fig.~\ref{fig:P_N2_mu1p0}.
The proposed model is in good agreement with the simulations.
 We observe that $N_2/\eta_0\dot{\gamma}$ is always negative, and is comparable to but smaller than $\eta_{\rm r}$.
 For volume fraction $\phi \le 0.45$, $\Pi/\eta_0\dot{\gamma}$ is smaller than $\eta_{\rm r}$. With increasing
 $\phi$ the particle pressure increases faster than the shear stress, and for $\phi$ approaching $\phi_{\rm J}(\mu)$,   
 $\Pi/\eta_0\dot{\gamma}$ becomes larger than the relative viscosity $\eta_{\rm r}$, as deduced in modeling based
  in part on particle migration data by Morris and Boulay~\cite{Morris_1999}. The experimental data by Boyer {\it et al.}~\cite{Boyer_2011}
  also show similar decrease in bulk friction coefficient as the jamming volume fraction is approached.
%
  
\begin{figure*}
\centering
\subfigure[]{
\includegraphics[width=.32\textwidth]{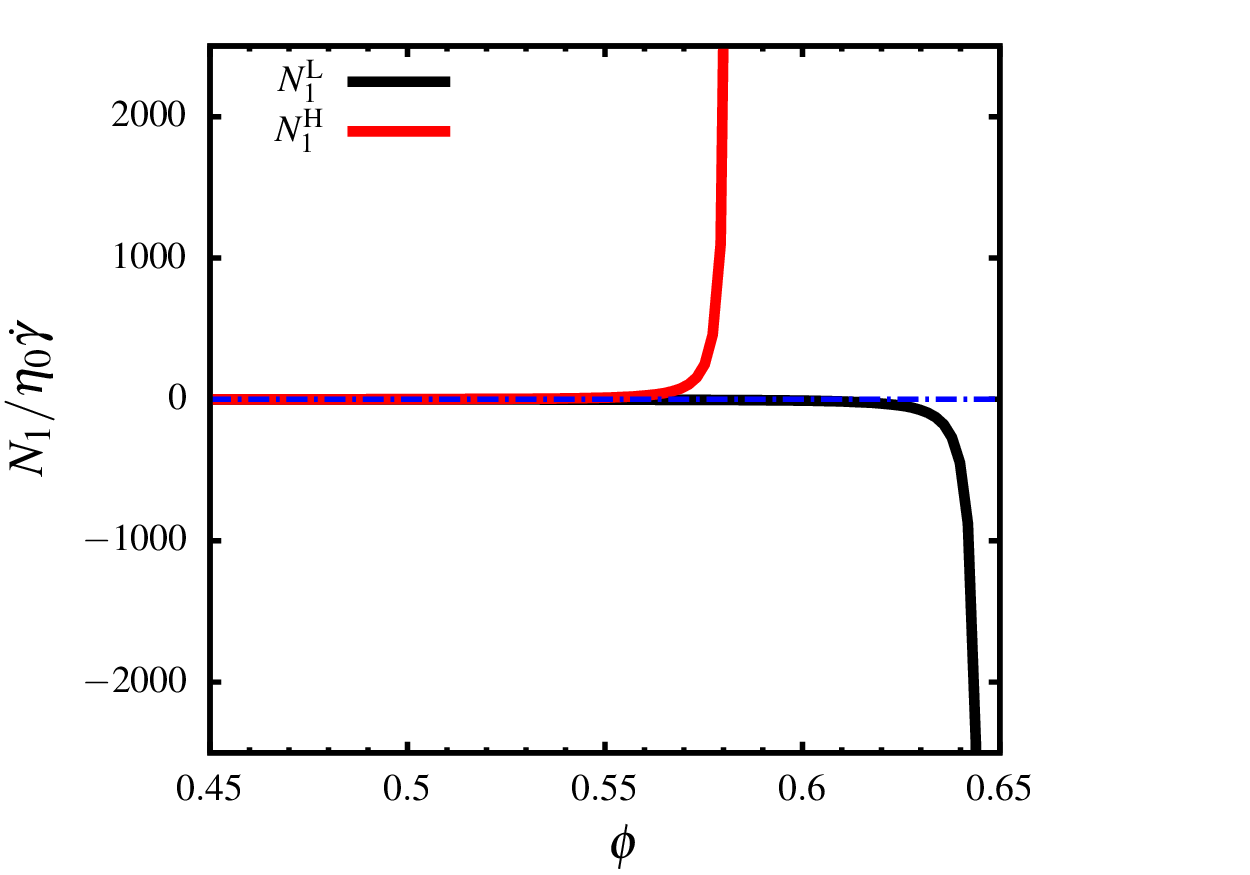}}
\subfigure[]{
\includegraphics[width=.32\textwidth]{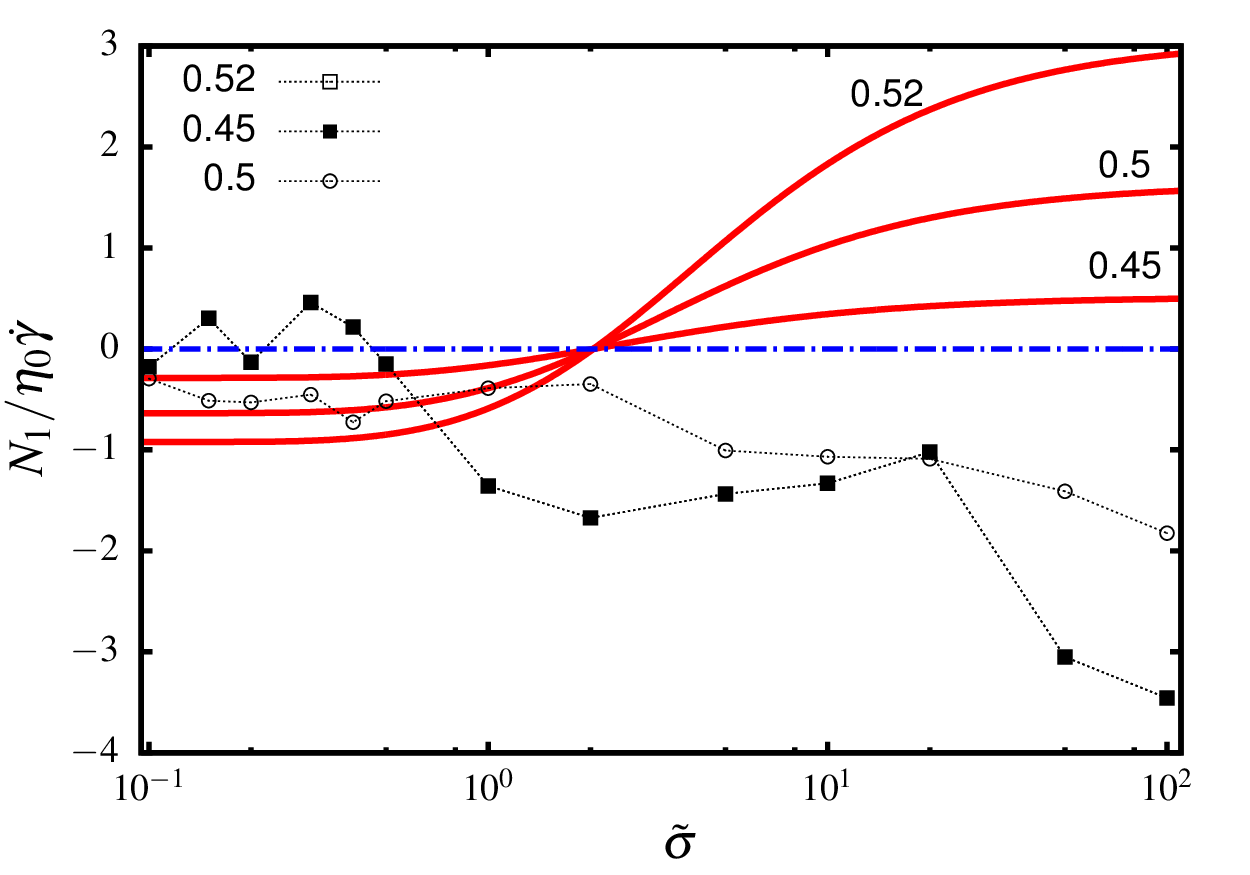}}
\subfigure[]{
\includegraphics[width=.32\textwidth]{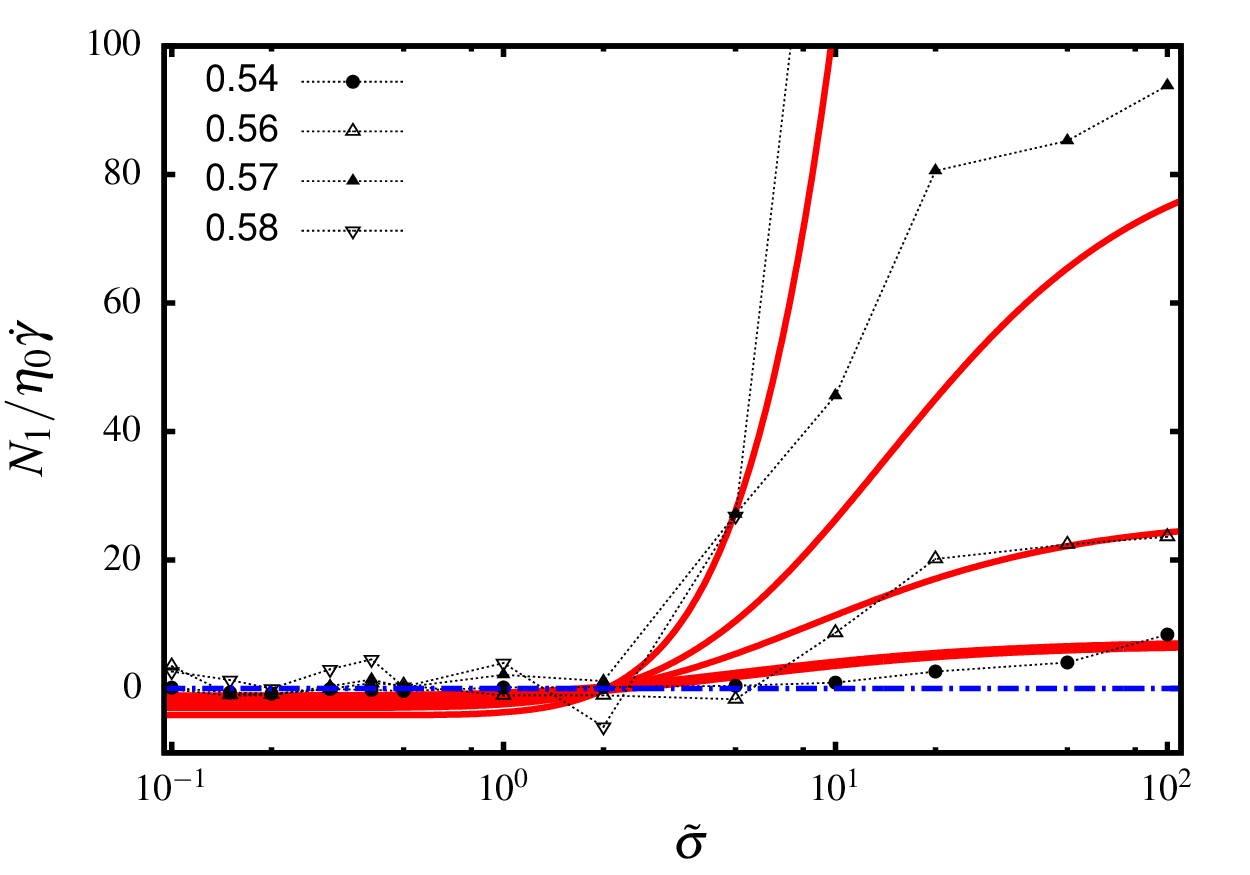}}
\caption{ (a) Divergence of the lubrication and frictional contributions of $N_1$. The lubrication part is negative and diverges at $\phi_{\rm J}^0$, 
while the frictional part is positive and diverges at $\phi_{\rm J}(\mu)$, and $K_1^{\rm 0} = 0.055$, $K_1(\mu) = 0.045$ 
(b) $N_1/\eta_0\dot{\gamma}(\phi,\tilde{\sigma})$ plotted as a function of $\tilde{\sigma} =\sigma/\sigma_0$ for volume fraction $\phi < 0.54$. 
(c) $N_1/\eta_0\dot{\gamma}(\phi,\tilde{\sigma})$ plotted as a function of $\tilde{\sigma} =\sigma/\sigma_0$ for high volume fraction $\phi \ge 0.54$.
Symbols and dashed lines indicate the simulation data while solid lines are predictions from~\eqref{eq:N1_str}.
}
\label{fig:N1_str}
\end{figure*}

 Finally, $N_1/\eta_0\dot{\gamma}$ from~\eqref{N1_phi_str_mu} is shown in Fig.~\ref{fig:N1_str}.
  Figure~\ref{fig:N1_str}a displays the divergences of $N_1/\eta_0\dot{\gamma}$ in stress-independent states, 
  where we choose the divergent volume fraction
  to be the same as that of $\eta_{\rm r}$ for $\mu=1$, where $K_1^{\rm 0} = 0.055$ and $K_1(\mu) = 0.045$ are used. 
  The predictions of the model for the stress-dependent $N_1/\eta_0\dot{\gamma}$ are plotted along with the simulation data in Figs.~\ref{fig:N1_str}b and ~\ref{fig:N1_str}c.
  The modelled $N_1/\eta_0\dot{\gamma}$ exhibits several features:
  for all volume fractions $\phi$, $N_1/\eta_0\dot{\gamma}$ is small and negative at small stress, and becomes increasingly negative with increasing stress, reaching a minimum
  value for $\tilde{\sigma} \approx 1$. The magnitude of this negative $N_1/\eta_0\dot{\gamma}$ becomes larger as $\phi$ is increased.
  At stress values $\tilde{\sigma}>1$, $N_1(\sigma)/\eta_0\dot{\gamma}$ tends toward positive values, crossing zero at $\tilde{\sigma}_{\rm p}$.
  The proposed model is in good agreement with the simulation data for the range of volume fraction $\phi \ge 0.54$,
  where the simulation data show positive $N_1$ at high stress. On the other hand, for volume fraction $\phi < 0.54$,
  the model does not agree with the data at high stress where simulations show negative $N_1$, while the model predicts
  $N_1$ to be positive.  
  The simulation data and model predictions for $N_1$ at larger $\phi$ agree in being negative at small stress
  (where lubrication films between most particles remain) and becoming 
  positive at large stress (where most of the contacts are frictional).  This also agrees in part with
  observations of Lootens {\it et al.}~\cite{Lootens_2005}, Dbouk {\it et al.}~\cite{Dbouk_2013}, and Royer {\it et al.}~\cite{Royer_2016},
  but not with the data of Cwalina and Wagner~\cite{Cwalina_2014}.
  However, at lower volume fractions, experiments which have 
  shown $N_1<0$ for all stresses~\cite{Lootens_2005,Cwalina_2014,Royer_2016} are in agreement with our simulations, but are not captured by the model.
  This suggests that there is a difference in microstructure between the lower and higher particle fractions such that behavior consistent with the lubricated regime is observed in the lower-$\phi$ suspension even when contacts are frictional. 
 
 \section{Discussion and conclusions}

The rheology of dense frictional suspensions determined by the extensive numerical simulations presented here displays continuous and discontinuous shear thickening and, at sufficiently large $\phi$, shear-induced jamming.  All of these behaviors are predicted by the model structure of Wyart and Cates \cite{Wyart_2014}.
We provide a thorough examination of 
the influence of the interparticle friction coefficient on the jamming fraction in the near-hard-sphere limit. 
We find that the approach to the jamming point of both the shear and normal stresses scale well with volume fraction as $(\phi_{\rm J}-\phi)^{-2}$
for either the lubricated or frictional case.
The comprehensive and coherent database from these simulations allows us to make detailed comparisons against a constitutive model  
incorporating stress-dependent frictional effects in dense  suspensions.

We find that a  model defined by three parameters --- solid volume fraction $\phi$, dimensionless stress $\tilde{\sigma} = \sigma/\sigma_0$
 and interparticle friction coefficient $\mu$ --- captures well the extreme rate-dependence of the rheology of these materials. 
 Here  $\sigma_0=F_0/6\pi a^2$ is a stress
 scale determined by a stabilizing repulsive force of magnitude $F_0$ at contact for particles of radius $a$.
  The central concept is that this stress scale divides the material response into low stress and high stress regions: 
  when the stress is small, $\tilde{\sigma} \ll 1$, particle surfaces remain separated by lubrication films and the viscosity and normal stresses are relatively small.
  When the stress overwhelms the repulsive force, i.e., when $\tilde{\sigma} \gg 1$, frictional contacts dominate and the rheological functions are much larger.
  This leads to two limiting jamming fractions: $\phi_{\rm J}^0$ in the frictionless states at low stress, and $\phi_{\rm J}({\mu})<\phi_{\rm J}^0$ at high stress. 
  A stress-dependent jamming fraction $\phi_{\rm m}(\tilde{\sigma},\mu)$  can be defined by interpolating between these two jamming fractions as a function of the applied stress, as shown by~\eqref{phi_str},
  in the manner proposed by Wyart and Cates~\cite{Wyart_2014} using the fraction of frictional contacts. The divergence of the stresses
  approaching the interpolated jamming fraction $\phi_{\rm m}$ follows the form of the two limits, with the stresses
  growing as  $(\phi_{\rm m} - \phi)^{-2}$.

  Based on these concepts, a constitutive model for dense frictional suspensions in steady simple shear flow has been proposed. 
  Comparison of the model predictions, e.g. relative viscosity $\eta_{\rm r}(\phi,\tilde{\sigma},\mu)$, agree well over the full range of parameters with the simulations 
  reported here. The overall behavior is described by a flow-state diagram given in Fig.~\ref{fig:phase_diagram}.
  This diagram, particularly in the $\phi$ - $\tilde{\sigma}$ plane, displays the various regions of material behavior obtainable at a fixed value of $\mu$.
  At smaller $\phi$, the material shear thickens continuously (CST), while above a critical solid fraction $\phi_{\rm C}$ the shear thickening becomes discontinuous.
  We find two regimes of DST:
(i) a pure DST regime between two {\it flowing} states for $\phi_{\rm C} < \phi < \phi_{\rm J}({\mu})$,
and
(ii) a DST-SJ regime where with increase of $\tilde{\sigma}$, DST gives way to shear-jamming for $\phi_{\rm J}({\mu}) < \phi < \phi_{\rm J}^0$.
In both the scenarios (i) and (ii), upon increase in $\tilde{\sigma}$ the suspension under shear goes through CST over a range of stress before entering DST.
With increasing $\mu$,  $\phi_{\rm J}({\mu})$ decreases, and as a consequence the range of $\phi$ over which shear jamming
is observed, i.e. $\phi_{\rm J}({\mu}) < \phi < \phi_{\rm J}^0$, increases. The range of volume fraction $\phi_{\rm C} < \phi < \phi_{\rm J}({\mu})$
 for which DST is observed also broadens with increase of $\mu$.
The onset stress to observe the shear-jammed state decreases with increase in $\phi$, and $\tilde{{\sigma}_{\rm J}} \rightarrow 0$
 as $\phi \rightarrow \phi_{\rm J}^0$. 

 Once the key parameters in the model are fitted the entire flow-state diagram can be constructed. To achieve the fitting, 
 a measure of the two jamming volume fractions ($\phi_{\rm J}^0$ and $\phi_{\rm J}({\mu})$) and a stress ramp $\eta_{\rm r}(\sigma)$
 at one volume fraction $\phi$ are required. The two jamming fractions can also be extracted from stress ramps
 at several $\phi$ provided these are sufficiently concentrated.

Finally, we expect that the formulation of the model itself should be robust to changes of particle properties such as 
polydispersity, particle shape, and other surface properties.
 However, the values of the parameters such as $\phi_{\rm J}$ are known to be sensitive to these details \citep{Hecke_2009, Liu_2010, Liu_2010a}.
 Extending this framework to Brownian \citep{Mari_2015} and cohesive suspensions \citep{Pednekar_2017}, where a strong shear thinning and yielding behavior
are observed, would be valuable.

\section{Acknowledgments}
Our code makes use of the CHOLMOD library by Tim Davis
(\url{http://faculty.cse.tamu.edu/davis/suitesparse.html}) for direct
Cholesky factorization of the sparse resistance matrix.
This work was supported, in part, under
National Science Foundation Grants CNS-0958379, CNS-
0855217, ACI-1126113 and the City University of New York
High Performance Computing Center at the College of Staten
Island.  JFM was supported by NSF 1605283.
\bibliography{dst}
\section{Supplementary information}

\subsection{Effect of friction on normal stresses}
\begin{figure*}
\centering
\subfigure[]{
\includegraphics[width=.32\textwidth]{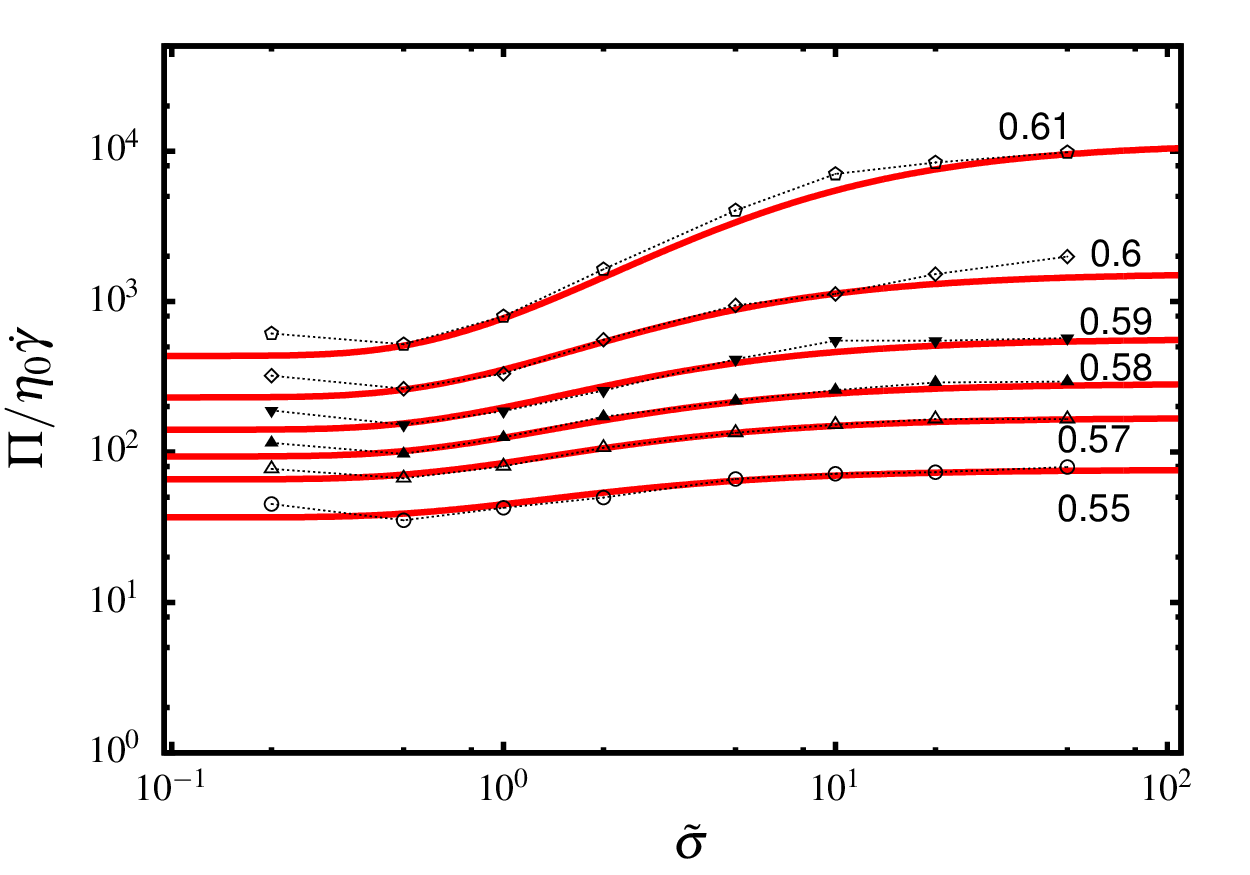}}
\subfigure[]{
\includegraphics[width=.32\textwidth]{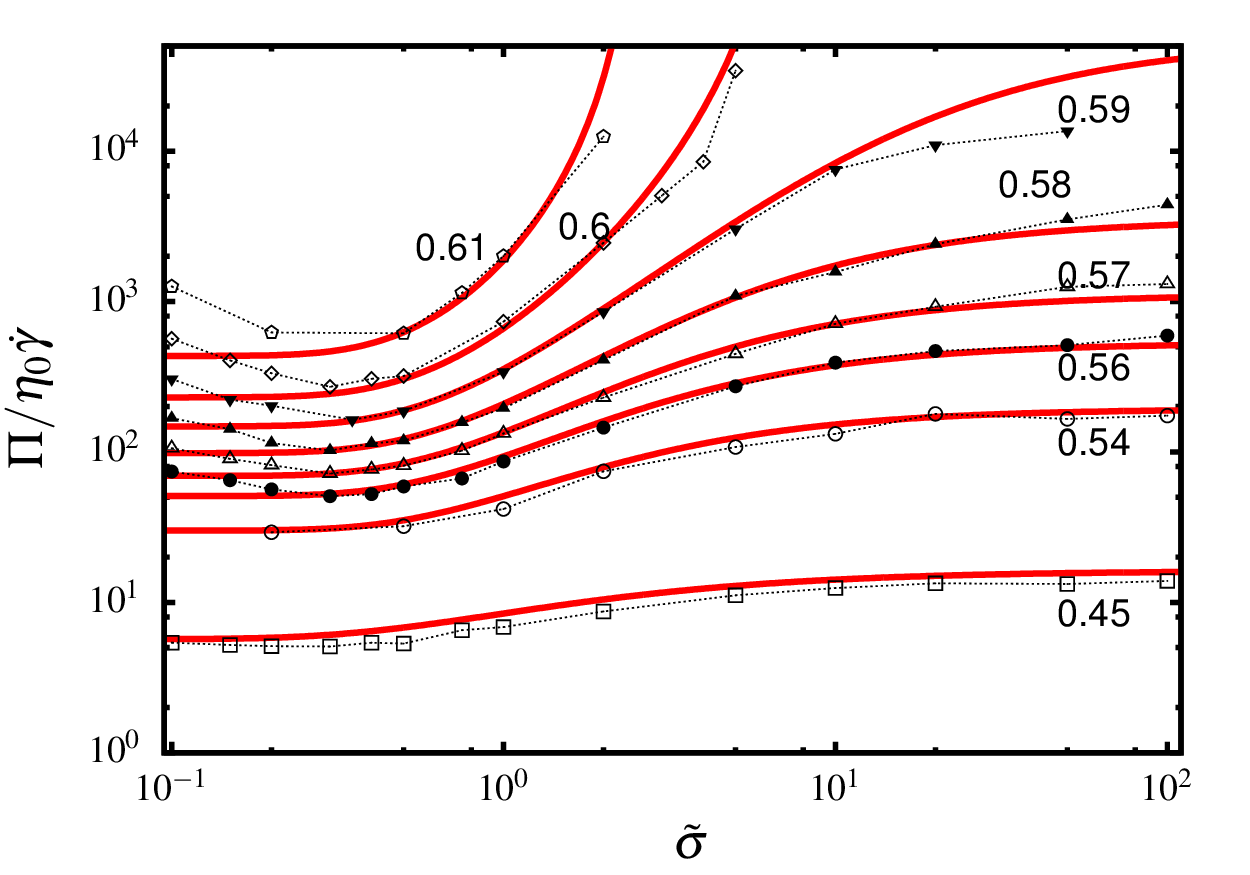}}
\subfigure[]{
\includegraphics[width=.32\textwidth]{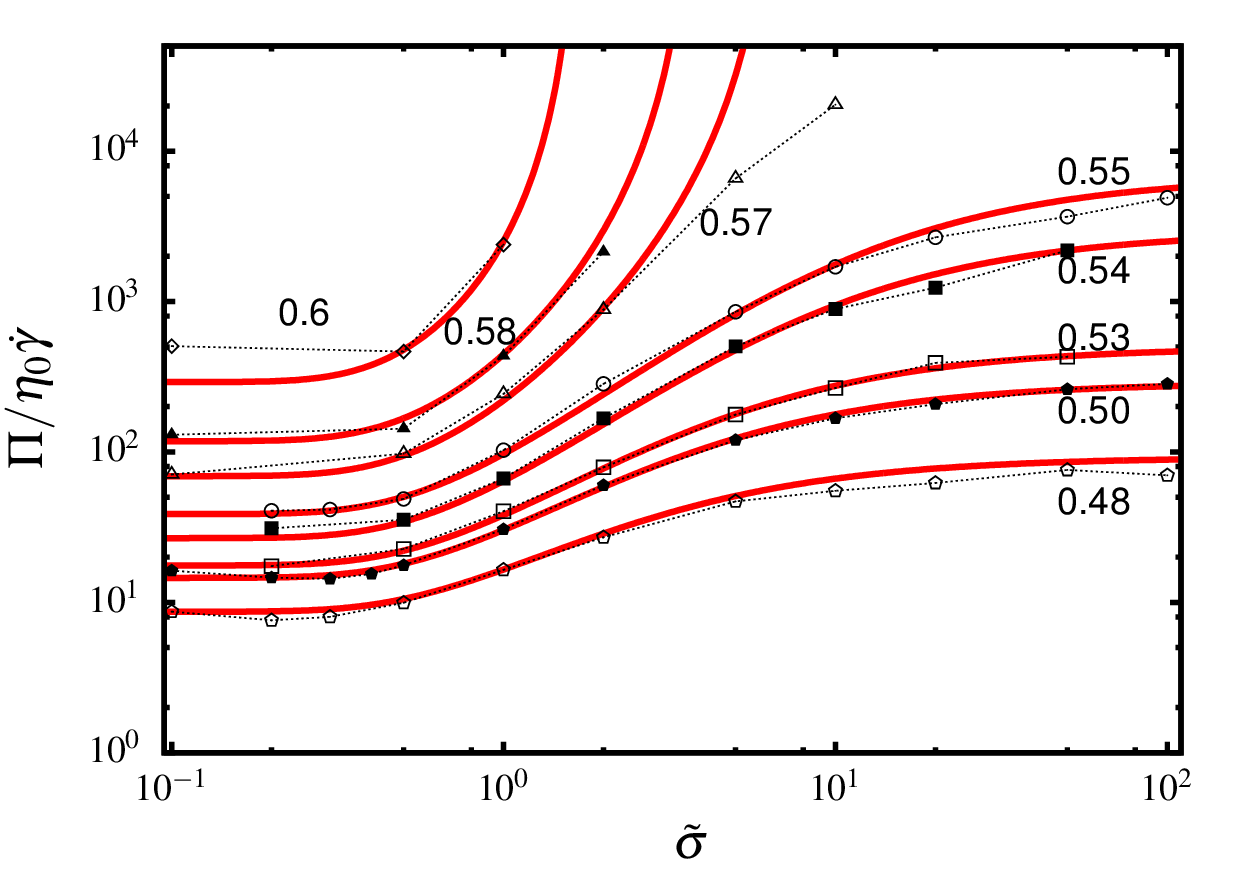}}
\caption{ Steady state particle pressure $\Pi/\eta_0\dot{\gamma}$ plotted against applied stress $\tilde{\sigma} = \sigma/\sigma_0$  for
$\mu = $ (a) 0.2, (b) 0.5, (c) 10 for several values of volume fractions as mentioned.
Symbols and dashed lines indicate the simulation data while solid lines are predictions from~\eqref{eq:P_str_phi_mu}.
}
\label{fig:SI_Press_fric_full_comp}
\end{figure*}

\begin{figure*}
\centering
\subfigure[]{
\includegraphics[width=.32\textwidth]{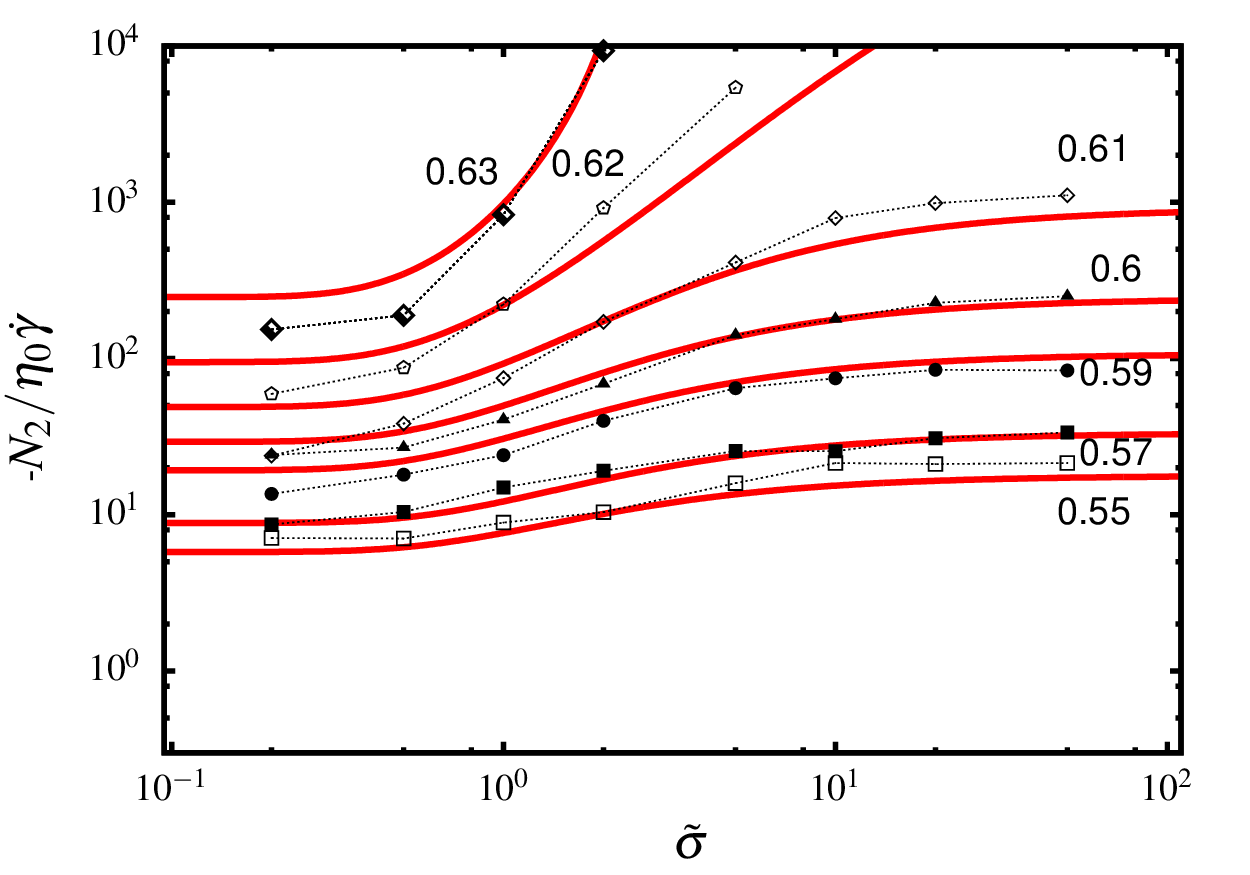}}
\subfigure[]{
\includegraphics[width=.32\textwidth]{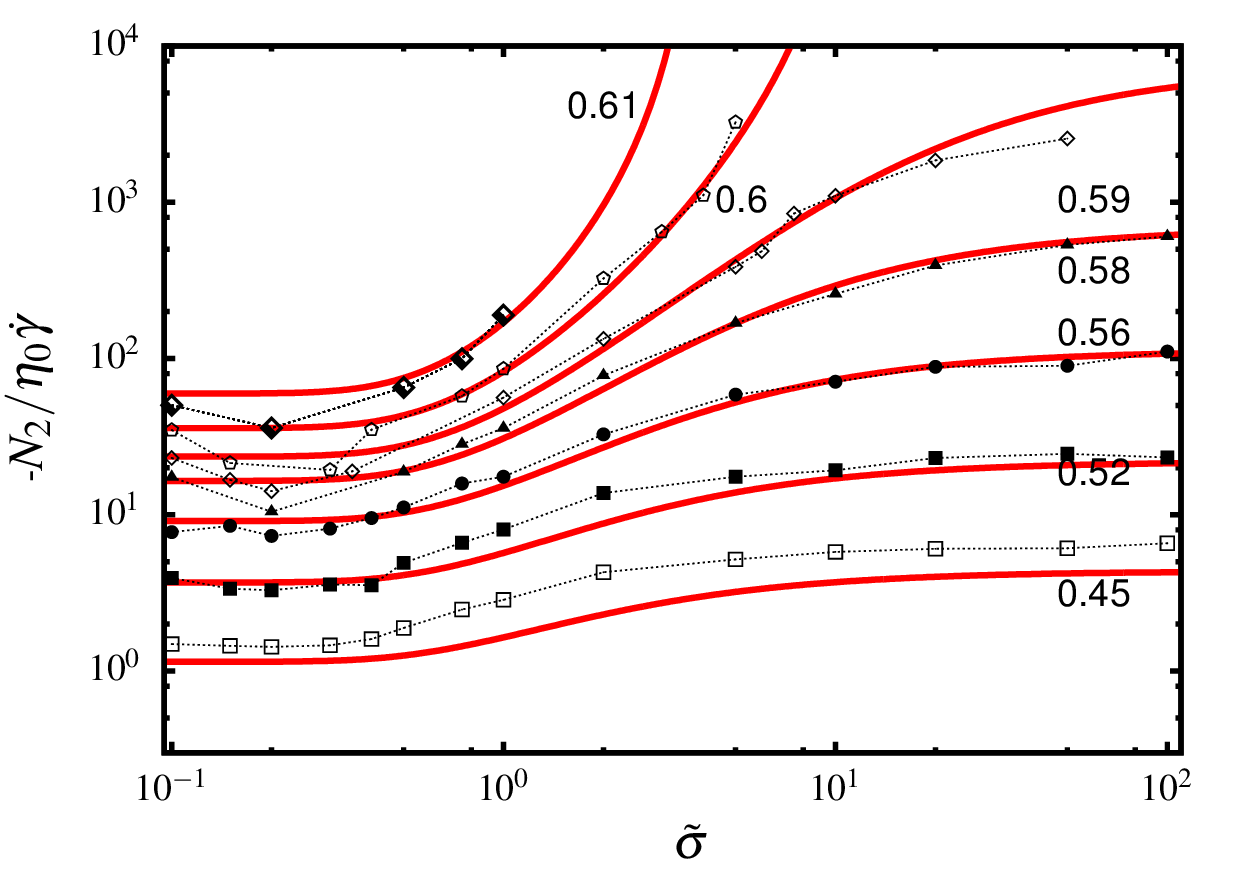}}
\subfigure[]{
\includegraphics[width=.32\textwidth]{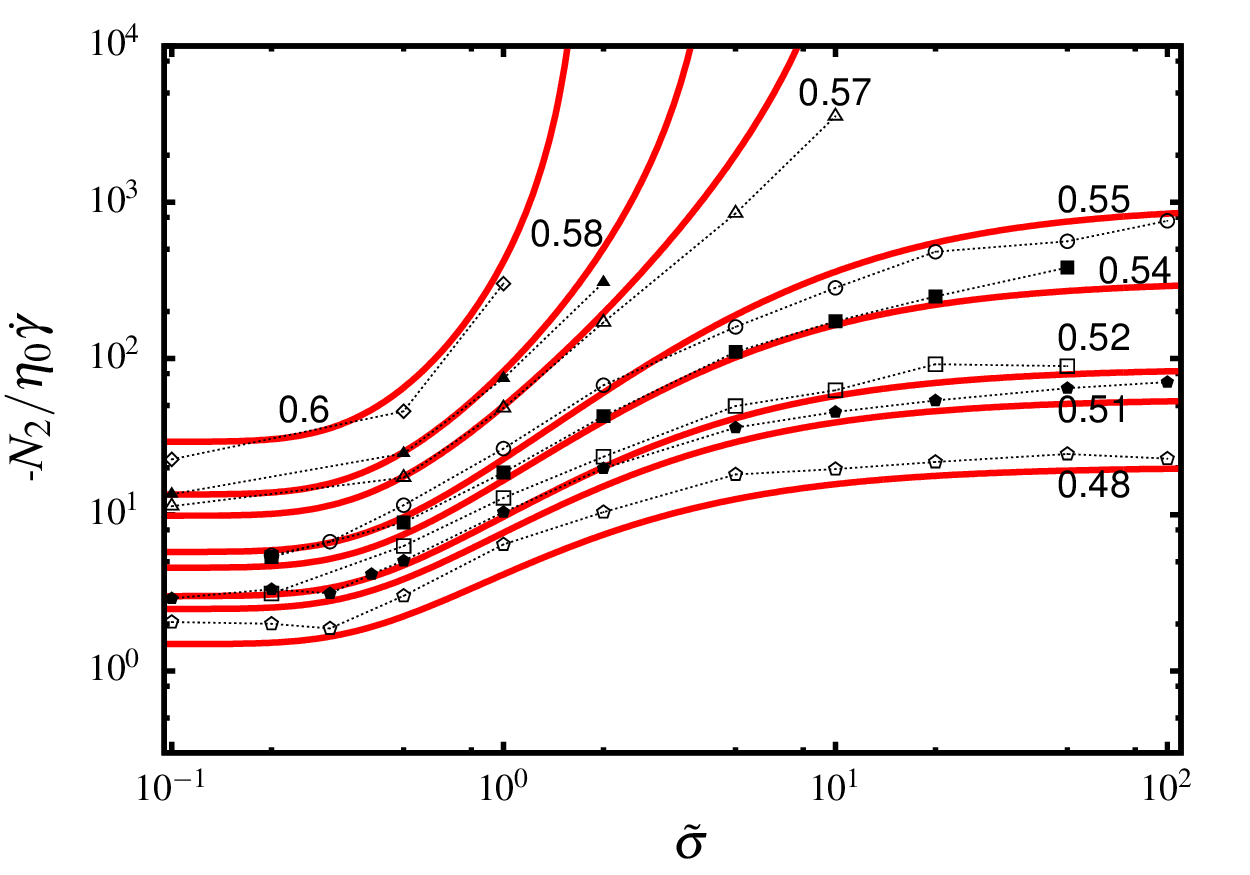}}
\caption{ Steady state second normal stress difference $N_2/\eta_0\dot{\gamma}$ plotted against applied stress $\tilde{\sigma} = \sigma/\sigma_0$  for
$\mu = $ (a) 0.2, (b) 0.5, (c) 10 for several values of volume fractions as mentioned.
Symbols and dashed lines indicate the simulation data while solid lines are predictions from~\eqref{eq:N2_str_phi_mu}.
}
\label{fig:SI_N2_fric_full_comp}
\end{figure*}

  Normal stresses ($\Pi/\eta_0\dot{\gamma}$ and $N_2/\eta_0\dot{\gamma}$) obtained from~\eqref{eq:eta_P_N2_phi_str}
  are presented in Figs.~\ref{fig:SI_Press_fric_full_comp} and ~\ref{fig:SI_N2_fric_full_comp}
  along with simulation data for different values of $\mu$.
 The model is in excellent agreement with the data. 
 For a given volume fraction $\phi$, both $\Pi/\eta_0\dot{\gamma}$ and $|N_2/\eta_0\dot{\gamma}|$ increase with $\mu$.

\end{document}